\documentclass[12pt]{article}
\usepackage{amssymb,amscd,array}
\usepackage[active]{srcltx}
\catcode `\@=11
\@addtoreset{equation}{section}

\newtheorem{thm}{Theorem}[section]

\newcommand{\be}{\begin{equation}}
\newcommand{\ee}{\end{equation}}
\newcommand{\bea}{\begin{eqnarray}}
\newcommand{\eea}{\end{eqnarray}}
\newcommand{\R}{\mathbb{R}}

\newcommand{\C}{\mathbb{C}}

\begin{document}
\begin{titlepage}

\begin{center}
{\bf \Large{Third Order Tree Contributions in the Causal Approach \\}}
\end{center}
\vskip 1.0truecm
\centerline{D. R. Grigore, 
\footnote{e-mail: grigore@theory.nipne.ro}}
\vskip5mm
\centerline{Department of Theoretical Physics,}
\centerline{Institute for Physics and Nuclear Engineering ``Horia Hulubei"}
\centerline{Bucharest-M\u agurele, P. O. Box MG 6, ROM\^ANIA}

\vskip 2cm
\bigskip \nopagebreak
\vskip 1cm
\begin{abstract}
\noindent
We consider the general framework of perturbative quantum field theory for the pure Yang-Mills model developped in \cite{wick+hopf}
and prove that the tree contributions do not give anomalies. We will provide a more general form of this gauge invariance property.
\end{abstract}

\end{titlepage}

\section{Introduction}

The most natural way to arrive at the Bogoliubov axioms of perturbative quantum field theory (pQFT) is by analogy with non-relativistic 
quantum mechanics \cite{Gl}, \cite{H}, \cite{D}, \cite{DF}: in this way one arrives naturally at Bogoliubov axioms 
\cite{BS}, \cite{EG}, \cite{Sc1}, \cite{Sc2}. We prefer the formulation from \cite{DF} and as presented in \cite{wick+hopf}; 
for every set of monomials 
$ 
A_{1}(x_{1}),\dots,A_{n}(x_{n}) 
$
in some jet variables (associated to some classical field theory) one associates the operator-valued distributions
$ 
T^{A_{1},\dots,A_{n}}(x_{1},\dots,x_{n})
$  
called chronological products; it will be convenient to use another notation: 
$ 
T(A_{1}(x_{1}),\dots,A_{n}(x_{n})). 
$ 

The Bogoliubov axioms express essentially some properties of the scattering matrix understood as a formal perturbation
series with the ``coefficients" the chronological products: 
\begin{itemize}
\item 
(skew)symmetry properties in the entries 
$ 
A_{1}(x_{1}),\dots,A_{n}(x_{n}) 
$;
\item
Poincar\'e invariance; 
\item
causality; 
\item
unitarity; 
\item
the ``initial condition" which says that
$
T(A(x)) 
$
is a Wick monomial.
\end{itemize}

So we need some basic notions on free fields and Wick monomials. One can supplement these axioms by requiring 
\begin{itemize}
\item 
power counting;
\item
Wick expansion property. 
\end{itemize}

It is a highly non-trivial problem to find solutions for the Bogoliubov axioms, even in the simplest case of a real scalar field. 
The simplest way is, in our opinion the procedure of Epstein and Glaser; it is a recursive construction for the basic objects
$ 
T(A_{1}(x_{1}),\dots,A_{n}(x_{n}))
$
and reduces the induction procedure to a distribution splitting of some distributions with causal support.  
In an equivalent way, one can reduce the induction procedure to the process of extension of distributions \cite{PS}. 

In the next Section we give some introductory facts concerning Wick products, Wick submonomials, Wick theorem and pure Yang-Mills
fields. The original results are given in Section \ref{tree-third} where we investigate chronological products of the form
$
T(A_{1}^{(2)}(x_{1}), A_{2}^{(2)}(x_{2}), A_{3}^{(1)}(x_{3})).
$
The tree contribution in the third order of the perturbation theory is a sum of such expressions . So third order gauge invariance 
in the tree sector must be
\bea
sT_{\rm tree}(A_{1}(x_{1}), A_{2}(x_{2}), A_{3}(x_{3})) = sT(A_{1}^{(2)}(x_{1}), A_{2}^{(2)}(x_{2}), A_{3}^{(1)}(x_{3}))
\nonumber\\
+sT(A_{1}^{(1)}(x_{1}), A_{2}^{(2)}(x_{2}), A_{3}^{(2)}(x_{3})) + sT(A_{1}^{(2)}(x_{1}), A_{2}^{(1)}(x_{2}), A_{3}^{(1)}(x_{3}))
\eea
(with apropiate Grassmann signs). We will prove that a stronger result is true, namely the every one of the third terms above is null. 

\newpage
\section{Perturbative Quantum Field Theory\label{pQFT}}
There are two main ingrediants in the contruction of a perturbative quantum field theory (pQFT): the construction of the Wick monomials 
and the Bogoliubov axioms. For a pQFT of Yang-Mills theories one needs one more ingrediant, namely the introduction of ghost fields and
gauge charge.

\subsection{Wick Products\label{wick prod}}

We consider a classical field theory on the Minkowski space
$
{\cal M} \simeq \R^{4}
$
(with variables
$
x^{\mu}, \mu = 0,\dots,3
$
and the metric $\eta$ with 
$
diag(\eta) = (1,-1,-1,-1)
$)
described by the Grassmann manifold 
$
\Xi_{0}
$
with variables
$
\xi_{a}, a \in {\cal A}
$
(here ${\cal A}$ is some index set) and the associated jet extension
$
J^{r}({\cal M}, \Xi_{0}),~r \geq 1
$
with variables 
$
x^{\mu},~\xi_{a;\mu_{1},\dots,\mu_{n}},~n = 0,\dots,r;
$
we denote generically by
$
\xi_{p}, p \in P
$
the variables corresponding to classical fields and their formal derivatives and by
$
\Xi_{r}
$
the linear space generated by them. The variables from
$
\Xi_{r}
$
generate the algebra
$
{\rm Alg}(\Xi_{r})
$
of polynomials.

To illustrate this, let us consider a real scalar field in Minkowski space ${\cal M}$. The first jet-bundle extension is
$$
J^{1}({\cal M}, \R) \simeq {\cal M} \times \R \times \R^{4}
$$
with coordinates 
$
(x^{\mu}, \phi, \phi_{\mu}),~\mu = 0,\dots,3.
$

If 
$
\varphi: \cal M \rightarrow \R
$
is a smooth function we can associate a new smooth function
$
j^{1}\varphi: {\cal M} \rightarrow J^{1}(\cal M, \R) 
$
according to 
$
j^{1}\varphi(x) = (x^{\mu}, \varphi(x), \partial_{\mu}\varphi(x)).
$

For higher order jet-bundle extensions we have to add new real variables
$
\phi_{\{\mu_{1},\dots,\mu_{r}\}}
$
considered completely symmetric in the indexes. For more complicated fields, one needs to add supplementary indexes to
the field i.e.
$
\phi \rightarrow \phi_{a}
$
and similarly for the derivatives. The index $a$ carries some finite dimensional representation of
$
SL(2,\C)
$
(Poincar\'e invariance) and, maybe a representation of other symmetry groups. 
In classical field theory the jet-bundle extensions
$
j^{r}\varphi(x)
$
do verify Euler-Lagrange equations. To write them we need the formal derivatives defined by
\be
d_{\nu}\phi_{\{\mu_{1},\dots,\mu_{r}\}} \equiv \phi_{\{\nu,\mu_{1},\dots,\mu_{r}\}}.
\ee

We suppose that in the algebra 
$
{\rm Alg}(\Xi_{r})
$
generated by the variables 
$
\xi_{p}
$
there is a natural conjugation
$
A \rightarrow A^{\dagger}.
$
If $A$ is some monomial in these variables, there is a canonical way to associate to $A$ a Wick 
monomial: we associate to every classical field
$
\xi_{a}, a \in {\cal A}
$
a quantum free field denoted by
$
\xi^{\rm quant}_{a}(x), a \in {\cal A}
$
and determined by the $2$-point function
\be
<\Omega, \xi^{\rm quant}_{a}(x), \xi^{\rm quant}_{b}(y) \Omega> = - i~D_{ab}^{(+)}(x - y)\times {\bf 1}.
\label{2-point}
\ee
Here 
\be
D_{ab}(x) = D_{ab}^{(+)}(x) + D_{ab}^{(-)}(x)
\ee
is the causal Pauli-Jordan distribution associated to the two fields; it is (up to some numerical factors) a polynomial
in the derivatives applied to the Pauli-Jordan distribution. We understand by 
$
D^{(\pm)}_{ab}(x)
$
the positive and negative parts of
$
D_{ab}(x)
$.
From (\ref{2-point}) we have
\be
[ \xi_{a}(x), \xi_{b}(y) ] = - i~ D_{ab}(x - y) \times {\bf 1} 
\ee
where by 
$
[\cdot, \cdot ]
$
we mean the graded commutator. 

The $n$-point functions for
$
n \geq 3
$
are obtained assuming that the truncated Wightman functions are null: see \cite{BLOT}, relations (8.74) and (8.75) and proposition 8.8
from there. The definition of these truncated Wightman functions involves the Fermi parities
$
|\xi_{p}|
$
of the fields
$
\xi_{p}, p \in P.
$

Afterwards we define
$$
\xi^{\rm quant}_{a;\mu_{1},\dots,\mu_{n}}(x) \equiv \partial_{\mu_{1}}\dots \partial_{\mu_{n}}\xi^{\rm quant}_{a}(x), a \in {\cal A}
$$
which amounts to
\be
[ \xi_{a;\mu_{1}\dots\mu_{m}}(x), \xi_{b;\nu_{1}\dots\nu_{n}}(y) ] =
(-1)^{n}~i~\partial_{\mu_{1}}\dots \partial_{\mu_{m}}\partial_{\nu_{1}}\dots \partial_{\nu_{n}}D_{ab}(x - y )\times {\bf 1}.
\label{2-point-der}
\ee
More sophisticated ways to define the free fields involve the GNS construction. 

The free quantum fields are generating a Fock space 
$
{\cal F}
$
in the sense of the Borchers algebra: formally it is generated by states of the form
$
\xi^{\rm quant}_{a_{1}}(x_{1})\dots \xi^{\rm quant}_{a_{n}}(x_{n})\Omega
$
where 
$
\Omega
$
the vacuum state.
The scalar product in this Fock space is constructed using the $n$-point distributions and we denote by
$
{\cal F}_{0} \subset {\cal F}
$
the algebraic Fock space.

One can prove that the quantum fields are free, i.e.
they verify some free field equation; in particular every field must verify Klein Gordon equation for some mass $m$
\be
(\square + m^{2})~\xi^{\rm quant}_{a}(x) = 0
\label{KG}
\ee
and it follows that in momentum space they must have the support on the hyperboloid of mass $m$. This means that 
they can be split in two parts
$
\xi^{\rm quant (\pm)}_{a}
$
with support on the upper (resp. lower) hyperboloid of mass $m$. We convene that 
$
\xi^{\rm quant (+)}_{a} 
$
resp.
$
\xi^{\rm quant (-)}_{a} 
$
correspond to the creation (resp. annihilation) part of the quantum field. The expressions
$
\xi^{\rm quant (+)}_{p} 
$
resp.
$
\xi^{\rm quant (-)}_{p} 
$
for a generic
$
\xi_{p},~ p \in P
$
are obtained in a natural way, applying partial derivatives. For a general discussion of this method of constructing free fields, 
see ref. \cite{BLOT} - especially prop. 8.8. We will follow essentially the presentation from \cite{wick+hopf}.
The Wick monomials are leaving invariant the algebraic Fock space.

\newpage
\subsection{Yang-Mills Fields\label{ym}}

First, we can generalize the preceding formalism to the case when some of the scalar fields
are odd Grassmann variables. One simply insert everywhere the Fermi sign. The next generalization is to arbitrary vector and spinorial
fields. If we consider for instance the Yang-Mills interaction Lagrangian corresponding to pure QCD then the jet variables 
$
\xi_{a}, a \in \Xi
$
are
$
(v^{\mu}_{a}, u_{a}, \tilde{u}_{a}),~a = 1,\dots,r
$
where 
$
v^{\mu}_{a}
$
are Grassmann even and 
$
u_{a}, \tilde{u}_{a}
$
are Grassmann odd variables. 

The interaction Lagrangian is determined by gauge invariance. Namely we define the {\it gauge charge} operator by
\be
d_{Q} v^{\mu}_{a} = i~d^{\mu}u_{a},\qquad
d_{Q} u_{a} = 0,\qquad
d_{Q} \tilde{u}_{a} = - i~d_{\mu}v^{\mu}_{a},~a = 1,\dots,r
\ee
where 
$
d^{\mu}
$
is the formal derivative. The gauge charge operator squares to zero:
\be
d_{Q}^{2} \simeq  0
\ee
where by
$
\simeq
$
we mean, modulo the equation of motion. Now we can define the interaction Lagrangian by the relative cohomology relation:
\be
d_{Q}T(x) \simeq {\rm total~divergence}.
\ee
If we eliminate the corresponding coboundaries, then a tri-linear Lorentz covariant 
expression is uniquely given by
\bea
T = f_{abc} \left( {1\over 2}~v_{a\mu}~v_{b\nu}~F_{c}^{\nu\mu}
+ u_{a}~v_{b}^{\mu}~d_{\mu}\tilde{u}_{c}\right)
\label{Tint}
\eea
where
\be
F^{\mu\nu}_{a} \equiv d^{\mu}v^{\nu}_{a} - d^{\nu}v^{\mu}_{a}, 
\quad \forall a = 1,\dots,r
\ee 
and 
$
f_{abc}
$
are real and completely anti-symmetric. (This is the tri-linear part of the usual QCD interaction Lagrangian from classical field theory.)

Then we define the associated Fock space by the non-zero $2$-point distributions are
\bea
<\Omega, v^{\mu}_{a}(x_{1}) v^{\nu}_{b}(x_{2})\Omega> = 
i~\eta^{\mu\nu}~\delta_{ab}~D_{0}^{(+)}(x_{1} - x_{2}),
\nonumber \\
<\Omega, u_{a}(x_{1}) \tilde{u}_{b}(x_{2})\Omega> = - i~\delta_{ab}~D_{0}^{(+)}(x_{1} - x_{2}),
\nonumber\\
<\Omega, \tilde{u}_{a}(x_{1}) u_{b}(x_{2})\Omega> = i~\delta_{ab}~D_{0}^{(+)}(x_{1} - x_{2}).
\label{2-massless-vector}
\eea
and construct the associated Wick monomials. Then the expression (\ref{Tint}) gives a Wick polynomial 
$
T^{\rm quant}
$
formally the same, but: 
(a) the jet variables must be replaced by the associated quantum fields; (b) the formal derivative 
$
d^{\mu}
$
goes in the true derivative in the coordinate space; (c) Wick ordering should be done to obtain well-defined operators. We also 
have an associated {\it gauge charge} operator in the Fock space given by
\bea
~[Q, v^{\mu}_{a}] = i~\partial^{\mu}u_{a},\qquad
\{ Q, u_{a} \} = 0,\qquad
\{Q, \tilde{u}_{a}\} = - i~\partial_{\mu}v^{\mu}_{a}
\nonumber \\
Q \Omega = 0.
\label{Q-vector-null}
\eea

Then it can be proved that
$
Q^{2} = 0
$
and
\be
~[Q, T^{\rm quant}(x) ] = {\rm total~divergence}
\label{gauge1}
\ee
where the equations of motion are automatically used because the quantum fields are on-shell.
From now on we abandon the super-script {\it quant} because it will be obvious from the context if we refer 
to the classical expression (\ref{Tint}) or to its quantum counterpart.

In (\ref{2-massless-vector}) we are using the Pauli-Jordan distribution
\be
D_{m}(x) = D_{m}^{(+)}(x) + D_{m}^{(-)}(x)
\ee
where
\be
D_{m}^{(\pm)}(x) =
\pm {i \over (2\pi)^{3}}~\int dp e^{- i p\cdot x} \theta(\pm p_{0}) \delta(p^{2} -
m^{2})
\ee
and
\be
D^{(-)}(x) = - D^{(+)}(- x).
\ee

We conclude our presentation with a generalization of (\ref{gauge1}). In fact, it can be proved that (\ref{gauge1}) implies
the existence of Wick polynomials
$
T^{\mu}
$
and
$
T^{\mu\nu}
$
such that we have:
\be
~[Q, T^{I} ] = i \partial_{\mu}T^{I\mu}
\label{gauge2}
\ee
for any multi-index $I$ with the convention
$
T^{\emptyset} \equiv T.
$
Explicitly:
\bea
T^{\mu} = f_{abc} \left( u_{a}~v_{b\nu}~F_{c}^{\nu\mu}
- {1\over 2}u_{a}~u_{b}~d^{\mu}\tilde{u}_{c}\right)
\label{Tmu-int}
\eea
and 
\bea
T^{\mu\nu} = {1\over 2}~f_{abc}~u_{a}~u_{b}~F_{c}^{\mu\nu}.
\label{Tmunu-int}
\eea

Finally we give the relation expressing gauge invariance in order $n$ of the perturbation theory. We define the operator 
$
\delta
$
on chronological products by:
\bea
\delta T(T^{I_{1}}(x_{1}),\dots,T^{I_{n}}(x_{n})) \equiv 
\sum_{m=1}^{n}~( -1)^{s_{m}}\partial_{\mu}^{m}T(T^{I_{1}}(x_{1}), \dots,T^{I_{m}\mu}(x_{m}),\dots,T^{I_{n}}(x_{n}))
\label{derT}
\eea
with
\be
s_{m} \equiv \sum_{p=1}^{m-1} |I_{p}|,
\ee
then we define the operator
\be
s \equiv d_{Q} - i \delta.
\label{s-n}
\ee

Gauge invariance in an arbitrary order is then expressed by
\be
sT(T^{I_{1}}(x_{1}),\dots,T^{I_{n}}(x_{n})) = 0.
\label{brst-n}
\ee
\newpage

\subsection{A More Precise Version of Wick Theorem\label{wick thm}}

For simplicity, we assume in this Section that the variables 
$
\xi_{a}
$
(see Subsection \ref{wick prod}) are commutative. We also use the summation convention over the dummy indices. In \cite{wick+hopf} we
have proved the following result.
\begin{thm}
Let us consider that
\be
A_{1} = {1\over 3!}~f_{pqr} \xi_{p}\xi_{q}\xi_{r},\qquad f_{pqr} = {\rm completely~ symmetric}
\ee
and
$
A_{2},\dots,A_{n}
$
are arbitrary. We define
\bea
T(A_{1}^{(1)}(x_{1}),\dots, A_{n}(x_{n})) \equiv T_{1}(A_{1}(x_{1}),\dots, A_{n}(x_{n})) =
\nonumber\\
{1 \over 2}~f_{pqr}~:\xi_{p}(x_{1})~T_{0}(\xi_{q}(x_{1})\xi_{r}(x_{1}),A_{2}(x_{2}), \dots, A_{n}(x_{n})):
\label{t31a}
\eea
\bea
T(A_{1}^{(2)}(x_{1}),\dots, A_{n}(x_{n})) \equiv T_{2}(A_{1}(x_{1}),\dots, A_{n}(x_{n})) =
\nonumber\\
{1 \over 2}~f_{pqr}~: \xi_{p}(x_{1})\xi_{q}(x_{1})~T(\xi_{r}(x_{1}),A_{2}(x_{2}), \dots, A_{n}(x_{n})):
\label{t32a}
\eea
\bea
T(A_{1}^{(3)}(x_{1}),\dots, A_{n}(x_{n})) \equiv T_{3}(A_{1}(x_{1}),\dots, A_{n}(x_{n})) =
\nonumber\\
{1 \over 3!}~f_{pqr}~: \xi_{p}(x_{1})\xi_{q}(x_{1})\xi_{r}(x_{1})~T(A_{2}(x_{2}), \dots, A_{n}(x_{n})):
\label{t33a}
\eea

and 
\be
T_{0} \equiv T - T_{1} - T_{2} - T_{3}.
\label{w31}
\ee
Then 
$
T_{0}
$
is of Wick type only in 
$
A_{2},\dots,A_{n}.
$
\end{thm}

We can iterate the arguments above in the entries 
$
A_{2},\dots,A_{n}
$
and obtain the following version of Wick theorem:
\be
T(A_{1}(x_{1}),\dots, A_{n}(x_{n})) = \sum T(A_{1}^{(k_{1})}(x_{1}),\dots, A_{n}^{(k_{n})}(x_{n}))
\label{w1-n}
\ee
where the sum runs over 
$
k_{1},\dots,k_{n} = 0,\dots,3
$
for 
$
A_{1},\dots,A_{n}
$
tri-linear. This formula can be written in a more transparent way if we use Hopf algebra notions - see \cite{wick+hopf}.
\newpage

\subsection{Wick submonomials in the pure Yang-Mills case\label{submonomials}}

We notice that in (\ref{Q-vector-null}) and in (\ref{gauge2}) we have a pattern of the type:
\be
d_{Q}A = {\rm total~divergence}.
\label{total-div}
\ee
This pattern remains true for Wick submonomials. We consider the expressions
(\ref{Tint}), (\ref{Tmu-int}) and (\ref{Tmunu-int}) from the pure Yang-Mills case and define:
\bea
B_{a\mu} \equiv \tilde{u}_{a,\mu} \cdot T = - f_{abc} u_{b}~v_{c\mu}
\nonumber\\
C_{a\mu} \equiv v_{a\mu} \cdot T = f_{abc} (v_{b}^{\nu}~F_{c\nu\mu} - u_{b}~\tilde{u}_{c,\mu})
\nonumber\\
D_{a} \equiv u_{a} \cdot T = f_{abc} v^{\mu}_{b}~\tilde{u}_{c,\mu}
\nonumber\\
E_{a\mu\nu} \equiv v_{a\mu,\nu} \cdot T = f_{abc} v_{b\mu}~v_{c\nu}
\nonumber\\
C_{a\nu\mu} \equiv v_{a\nu} \cdot T_{\mu} = - f_{abc} u_{b}~F_{c\nu\mu}.
\label{sub1}
\eea
We also have
\bea
u_{a} \cdot T = - C_{a\mu}
\nonumber\\
v_{a\rho,\sigma} \cdot T_{\mu} = \eta_{\mu\sigma}~B_{a\rho} -  \eta_{\mu\rho}~B_{a\sigma}
\nonumber\\
u_{a} \cdot T_{\mu\nu} = - C_{a\mu\nu}
\label{sub2}
\eea
If we define
\be
B_{a} \equiv {1 \over 2}~f_{abc}~u_{b}~u_{c}
\label{B}
\ee
we also have
\bea
\tilde{u}_{a,\nu} \cdot T_{\mu} = \eta_{\mu\nu}~B_{a}
\nonumber\\
v_{a\rho,\sigma} \cdot T_{\mu\nu} = (\eta_{\mu\sigma}~\eta_{\nu\rho} - \eta_{\nu\sigma}~\eta_{\mu\rho})~B_{a}.
\label{sub3}
\eea

Then we try to extend the structure (\ref{total-div}) to the Wick submonomials defined above. We have:
\bea
d_{Q} B_{a}^{\mu} = i d^{\mu}B_{a}
\nonumber\\
d_{Q} C_{a}^{\mu} = i d_{\nu}C_{a}^{\mu\nu}
\nonumber\\
d_{Q} D_{a} = - i d_{\mu}C_{a}^{\mu}
\nonumber\\
d_{Q} E_{a}^{\mu\nu} = i (d^{\nu}B_{a}^{\mu} -  d^{\mu}B_{a}^{\nu} + C_{a}^{\mu\nu}) 
\nonumber\\
d_{Q} B_{a} = 0
\nonumber\\
d_{Q} C_{a}^{\mu\nu} = 0. 
\label{dQB}
\eea
So we see that the patern (\ref{total-div}) is broken only for 
$
E_{a}^{\mu\nu}.
$
We fix this in the following way. We have the formal derivative
\be
\delta A \equiv d_{\mu}A^{\mu}
\ee
used in the definition of gauge invariance (\ref{derT}) + (\ref{s-n}); we also define the derivative
$
\delta^{\prime}
$
by
\be
\delta^{\prime}E_{a}^{\mu\nu} = C_{a}^{\mu\nu}
\label{dprime}
\ee
and $0$ for the other Wick submonomials (\ref{sub1}) and (\ref{B}). Finally
\be
s \equiv d_{Q} - i\delta, \qquad s^{\prime} \equiv s - i\delta^{\prime} = d_{Q} - i(\delta +\delta^{\prime}).
\ee
Then we have the structure
\be
s^{\prime} A = 0
\ee
for all expressions
$
A = T^{I}, 
$
$
B_{a\mu}, C_{a\mu},
$
etc. and also for the basic jet variables
$
v_{a\mu}, u_{a}, \tilde{u}_{a}.
$

\newpage
\subsection{Second Order Gauge Invariance. Tree Contributions\label{tree}}
We first have:
\bea
D(v^{\mu}_{a}(x_{1}), v^{\nu}_{b}(x_{2})) \equiv [ v^{\mu}_{a}(x_{1}), v^{\nu}_{b}(x_{2}) ] 
= i~\eta^{\mu\nu}~\delta_{ab}~D_{0}(x_{1} - x_{2}),
\nonumber \\
D(u_{a}(x_{1}), \tilde{u}_{b}(x_{2})) \equiv [ u_{a}(x_{1}), \tilde{u}_{b}(x_{2}) ] = - i~\delta_{ab}~D_{0}(x_{1} - x_{2}),
\nonumber\\
D(\tilde{u}_{a}(x_{1}), u_{b}(x_{2})) \equiv [ \tilde{u}_{a}(x_{1}), u_{b}(x_{2}) ] = i~\delta_{ab}~D_{0}(x_{1} - x_{2}).
\label{comm-2-massless-vector}
\eea
where in the left hand side we have the graded commutator. The causal splitting 
$
D = D^{\rm adv} - D^{\rm ret}
$
is unique because the degree of singularity of
$
D_{0}
$
is 
$
\omega = - 2
$ 
and we obtain 
\bea
T(v^{\mu}_{a}(x_{1})^{(0)}, v^{\nu}_{b}(x_{2})^{(0)}) = i~\eta^{\mu\nu}~\delta_{ab}~D^{F}_{0}(x_{1} - x_{2}),
\nonumber \\
T( u_{a}(x_{1})^{(0)}, \tilde{u}_{b}(x_{2})^{(0)}) = - i~\delta_{ab}~D^{F}_{0}(x_{1} - x_{2}),
\nonumber\\
T( \tilde{u}_{a}(x_{1})^{(0)}, u_{b}(x_{2})^{(0)}) = i~\delta_{ab}~D^{F}_{0}(x_{1} - x_{2}).
\label{chr-2-massless-vector}
\eea
From the previous relations we also have uniquely:
\bea
T(\xi^{(0)}_{a,\mu}(x_{1}), \xi^{(0)}_{b}(x_{2})) = \partial_{\mu}^{1}D^{F}(\xi_{a}(x_{1}), \xi_{b}(x_{2}))
\nonumber\\
T(\xi^{(0)}_{a}(x_{1}), \xi^{(0)}_{b,\nu}(x_{2})) =  \partial_{\nu}^{2}D^{F}(\xi_{a}(x_{1}), \xi_{b}(x_{2})).
\eea
However the causal splitting of 
$
T(\xi_{a,\mu}(x_{1})^{(0)}, \xi_{b,\nu}(x_{2})^{(0)})
$
is not unique because the distribution has the degree of singularity 
$
\omega = 0.
$
This was noticed for the first time in \cite{ASD} and \cite{DKS}. A possible choice is the {\it canonical} splitting, 
following from (\ref{2-point-der}):
\be
T(\xi^{(0)}_{a,\mu}(x_{1}), \xi^{(0)}_{b,\nu}(x_{2})) = 
i \partial_{\mu}^{1}\partial_{\nu}^{2}T(\xi_{a}(x_{1})^{(0)}, \xi_{b}(x_{2})^{(0)})
\label{canonical}
\ee

From these formulas we can determine now if gauge invariance is true; in fact, we have anomalies, as it is well known, but
they can be eliminated according to:
\begin{thm}
We have
\be
sT(T^{I_{1}}(x_{1}), T^{I_{2}}(x_{2})) = s^{\prime}T(T^{I_{1}}(x_{1})^{(2)}, T^{I_{2}}(x_{2})^{(2)})
\ee
and the anomalies given in the previous theorem can be eliminated if and only if the constants
$
f_{abc}
$
verify the Jacobi identity
\be
f_{eab}~f_{ecd} + f_{ebc}~f_{ead} + f_{eca}~f_{ebd}  = 0
\label{Jacobi}
\ee
using the finite renormalizations:
\be
T(A_{1}(x_{1}), A_{2}(x_{2})) \rightarrow T^{\rm ren}(A_{1}(x_{1}), A_{2}(x_{2})) 
= T(A_{1}(x_{1}), A_{2}(x_{2})) + \delta (x_{1} - x_{2})~N(A_{1},A_{2})(x_{2})
\label{finiteN}
\ee
where
\bea
N(T,T) \equiv {i \over 2}~E_{a}^{\mu\nu}~E_{a\mu\nu}
\nonumber\\
N(T^{\mu},T) = N(T, T^{\mu}) \equiv - i~B_{a\nu}~E_{a}^{\mu\nu}
\nonumber\\
N(T^{\mu},T^{\nu}) \equiv i~B_{a}^{\mu}~B_{a}^{\nu}
\nonumber\\
N(T^{\mu\nu},T) = N(T, T^{\mu\nu}) \equiv  - i~B_{a}~E_{a}^{\mu\nu}
\nonumber\\
N(T^{\mu\nu},T^{\rho}) = N(T^{\rho}, T^{\mu\nu}) = 0
\nonumber \\
N(T^{\mu\nu},T^{\rho\sigma}) = 0
\label{N-TT}
\eea
The previous finite renormalizations can be obtained performing the finite renormalization
\be
N(v_{a\mu,\nu},v_{b\rho,\sigma}) = {i \over 2} \eta_{\mu\rho}~\eta_{\nu\sigma}~\delta_{ab}.
\label{vv}
\ee
\label{ren-TT}
\end{thm}
\newpage
\section{Tree Contributions in the Third Order\label{tree-third}}

If we apply (\ref{w1-n}) in the third order of the perturbation theory we obtain the following tree contributions
\bea
T(T^{(2)}(x_{1}),T^{(2)}(x_{2}),T^{(1)}(x_{3})) = 
\nonumber\\
T(v_{a_{1}\mu_{1}}^{(0)}(x_{1}), v_{a_{2}\mu_{2}}^{(0)}(x_{2}), C_{a_{3}}^{\mu_{3}(0)}(x_{3}))~
C_{a_{1}}^{\mu_{1}}(x_{1})~C_{a_{2}}^{\mu_{2}}(x_{2})~v_{a_{3}\mu_{3}}(x_{3})
\nonumber\\
+ {1\over 2}~T(v_{a_{1}\mu_{1}}^{(0)}(x_{1}), v_{a_{2}\mu_{2}}^{(0)}(x_{2}), E_{a_{3}}^{\rho\sigma(0)}(x_{3}))~
C_{a_{1}}^{\mu_{1}}(x_{1})~C_{a_{2}}^{\mu_{2}}(x_{2})~F_{a_{3}\sigma\rho}(x_{3})
\nonumber\\
+ {1\over 4}~T(F_{a_{1}\sigma_{1}\rho_{1}}^{(0)}(x_{1}), F_{a_{2}\sigma_{2}\rho_{2}}^{(0)}(x_{2}), 
C_{a_{3}}^{\mu_{3}(0)}(x_{3}))~
E_{a_{1}}^{\rho_{1}\sigma_{1}}(x_{1})~E_{a_{2}}^{\rho_{2}\sigma_{2}}(x_{2})~v_{a_{3}\mu_{3}}(x_{3})
\nonumber\\
+ {1\over 8}~T(F_{a_{1}\sigma_{1}\rho_{1}}^{(0)}(x_{1}), F_{a_{2}\sigma_{2}\rho_{2}}^{(0)}(x_{2}), 
E_{a_{3}}^{\rho_{3}\sigma_{3}(0)}(x_{3}))~
E_{a_{1}}^{\rho_{1}\sigma_{1}}(x_{1})~E_{a_{2}}^{\rho_{2}\sigma_{2}}(x_{2})~F_{a_{3}\sigma_{3}\rho_{3}}(x_{3})
\nonumber\\
+ {1\over 2}~T(v_{a_{1}\nu}^{(0)}(x_{1}), F_{a_{2}\sigma\rho}^{(0)}(x_{2}), C_{a_{3}}^{\mu(0)}(x_{3}))~
C_{a_{1}}^{\nu}(x_{1})~E_{a_{2}}^{\rho\sigma}(x_{2})~v_{a_{3}\mu}(x_{3}) + (x_{1} \leftrightarrow x_{2})
\nonumber\\
+ {1\over 4}~T(v_{a_{1}\lambda}^{(0)}(x_{1}), F_{a_{2}\sigma\rho}^{(0)}(x_{2}), E_{a_{3}}^{\mu\nu(0)}(x_{3}))~
C_{a_{1}}^{\lambda}(x_{1})~E_{a_{2}}^{\rho\sigma}(x_{2})~F_{a_{3}\nu\mu}(x_{3}) + (x_{1} \leftrightarrow x_{2})
\nonumber\\
+ T(v_{a_{1}\rho}^{(0)}(x_{1}), u_{a_{2}}^{(0)}(x_{2}), D_{a_{3}}^{(0)}(x_{3}))~
C_{a_{1}}^{\rho}(x_{1})~D_{a_{2}}(x_{2})~u_{a_{3}}(x_{3}) + (x_{1} \leftrightarrow x_{2})
\nonumber\\
+ \partial_{\nu}^{2}T(v_{a_{1}\rho}^{(0)}(x_{1}), \tilde{u}_{a_{2}}^{(0)}(x_{2}), B_{a_{3}}^{\mu(0)}(x_{3}))~
C_{a_{1}}^{\rho}(x_{1})~B_{a_{2}}^{\nu}(x_{2})~\partial_{\mu}\tilde{u}_{a_{3}}(x_{3}) + (x_{1} \leftrightarrow x_{2})
\nonumber\\
+ {1\over 2}~T(F_{a_{1}\sigma\rho}^{(0)}(x_{1}), u_{a_{2}}^{(0)}(x_{2}), D_{a_{3}}^{(0)}(x_{3}))~
E_{a_{1}}^{\rho\sigma}(x_{1})~D_{a_{2}}(x_{2})~u_{a_{3}}(x_{3}) + (x_{1} \leftrightarrow x_{2})
\nonumber\\
+ {1\over 2}~\partial_{\nu}^{2}
T(F_{a_{1}\sigma\rho}^{(0)}(x_{1}), \tilde{u}_{a_{2}}^{(0)}(x_{2}), B_{a_{3}}^{\mu(0)}(x_{3}))~
E_{a_{1}}^{\rho\sigma}(x_{1})~B_{a_{2}}^{\nu}(x_{2})~\partial_{\mu}\tilde{u}_{a_{3}}(x_{3}) 
+ (x_{1} \leftrightarrow x_{2})
\nonumber\\
- \partial_{\nu}^{2}T(u_{a_{1}}^{(0)}(x_{1}), \tilde{u}_{a_{2}}^{(0)}(x_{2}), C_{a_{3}}^{\mu(0)}(x_{3}))~
D_{a_{1}}(x_{1})~B_{a_{2}}^{\nu}(x_{2})~v_{a_{3}\mu}(x_{3}) + (x_{1} \leftrightarrow x_{2})
\label{2,2,1}
\eea
\bea
T(T^{(2)}(x_{1}),T^{(2)}(x_{2}),T^{\mu(1)}(x_{3})) = 
\nonumber\\
T(v_{a_{1}\mu_{1}}^{(0)}(x_{1}), v_{a_{2}\mu_{2}}^{(0)}(x_{2}), C_{a_{3}}^{\mu(0)}(x_{3}))~
C_{a_{1}}^{\mu_{1}}(x_{1})~C_{a_{2}}^{\mu_{2}}(x_{2})~u_{a_{3}}(x_{3})
\nonumber\\
+ {1\over 4}~T(F_{a_{1}\sigma_{1}\rho_{1}}^{(0)}(x_{1}), F_{a_{2}\sigma_{2}\rho_{2}}^{(0)}(x_{2}), 
C_{a_{3}}^{\mu(0)}(x_{3}))~
E_{a_{1}}^{\rho_{1}\sigma_{1}}(x_{1})~E_{a_{2}}^{\rho_{2}\sigma_{2}}(x_{2})~u_{a_{3}}(x_{3})
\nonumber\\
+ \partial_{\mu_{1}}^{1}\partial_{\mu_{2}}^{2}
T(\tilde{u}_{a_{1}}^{(0)}(x_{1}), \tilde{u}_{a_{2}}^{(0)}(x_{2}),B_{a_{3}}^{(0)}(x_{3}))~
B_{a_{1}}^{\mu_{1}}(x_{1})~B_{a_{2}}^{\mu_{2}}(x_{2})~\partial^{\mu}\tilde{u}_{a_{3}}(x_{3})
\nonumber\\
+ {1\over 2}~T(v_{a_{1}\nu}^{(0)}(x_{1}), F_{a_{2}\sigma\rho}^{(0)}(x_{2}), C_{a_{3}}^{\mu(0)}(x_{3}))~
C_{a_{1}}^{\nu}(x_{1})~E_{a_{2}}^{\rho\sigma}(x_{2})~u_{a_{3}}(x_{3}) + (x_{1} \leftrightarrow x_{2})
\nonumber\\
- \partial_{\lambda}^{2}T(v_{a_{1}\rho}^{(0)}(x_{1}), \tilde{u}_{a_{2}}^{(0)}(x_{2}), C_{a_{3}}^{\nu\mu(0)}(x_{3}))~
C_{a_{1}}^{\rho}(x_{1})~B_{a_{2}}^{\lambda}(x_{2})~v_{a_{3}\nu}(x_{3}) + (x_{1} \leftrightarrow x_{2})
\nonumber\\
- \partial_{\sigma}^{2}T(v_{a_{1}\rho}^{(0)}(x_{1}), \tilde{u}_{a_{2}}^{(0)}(x_{2}), B_{a_{3}\nu}^{(0)}(x_{3}))~
C_{a_{1}}^{\rho}(x_{1})~B_{a_{2}}^{\sigma}(x_{2})~F_{a_{3}}^{\mu\nu}(x_{3}) + (x_{1} \leftrightarrow x_{2})
\nonumber\\
- {1\over 2}~\partial_{\lambda}^{2}
T(F_{a_{1}\sigma\rho}^{(0)}(x_{1}), \tilde{u}_{a_{2}}^{(0)}(x_{2}), C_{a_{3}}^{\nu\mu(0)}(x_{3}))~
E_{a_{1}}^{\rho\sigma}(x_{1})~B_{a_{2}}^{\lambda}(x_{2})~v_{a_{3}\nu}(x_{3}) + (x_{1} \leftrightarrow x_{2})
\nonumber\\
- {1\over 2}~\partial_{\lambda}^{2}
T(F_{a_{1}\sigma\rho}^{(0)}(x_{1}), \tilde{u}_{a_{2}}^{(0)}(x_{2}), B_{a_{3}\nu}^{(0)}(x_{3}))~
E_{a_{1}}^{\rho\sigma}(x_{1})~B_{a_{2}}^{\lambda}(x_{2})~F_{a_{3}}^{\mu\nu}(x_{3}) + (x_{1} \leftrightarrow x_{2})
\nonumber\\
- \partial_{\nu}^{2}T(u_{a_{1}}^{(0)}(x_{1}), \tilde{u}_{a_{2}}^{(0)}(x_{2}), C_{a_{3}}^{\mu(0)}(x_{3}))~
D_{a_{1}}(x_{1})~B_{a_{2}}^{\nu}(x_{2})~u_{a_{3}}(x_{3}) + (x_{1} \leftrightarrow x_{2})
\label{2,2,1mu}
\eea
\newpage
\bea
T(T^{(1)}(x_{1}),T^{(2)}(x_{2}),T^{\mu(2)}(x_{3})) = 
\nonumber\\
T(C_{a_{1}}^{\rho(0)}(x_{1}), v_{a_{2}\sigma}^{(0)}(x_{2}), v_{a_{3}\nu}^{(0)}(x_{3}))~
v_{a_{1}\rho}(x_{1})~C_{a_{2}}^{\sigma}(x_{2})~C_{a_{3}}^{\nu\mu}(x_{3})
\nonumber\\
- T(C_{a_{1}}^{\rho(0)}(x_{1}), v_{a_{2}\sigma}^{(0)}(x_{2}), F_{a_{3}}^{\nu\mu(0)}(x_{3}))~
v_{a_{1}\rho}(x_{1})~C_{a_{2}}^{\sigma}(x_{2})~B_{a_{3}\nu}(x_{3})
\nonumber\\
+ {1\over 2}~T(C_{a_{1}}^{\lambda(0)}(x_{1}), F_{a_{2}\sigma\rho}^{(0)}(x_{2}), v_{a_{3}\nu}^{(0)}(x_{3}))~
v_{a_{1}\lambda}(x_{1})~E_{a_{2}}^{\rho\sigma}(x_{2})~C_{a_{3}}^{\nu\mu}(x_{3})
\nonumber\\
- {1\over 2}~T(C_{a_{1}}^{\lambda(0)}(x_{1}), F_{a_{2}\sigma\rho}^{(0)}(x_{2}), F_{a_{3}}^{\nu\mu(0)}(x_{3}))~
v_{a_{1}\lambda}(x_{1})~E_{a_{2}}^{\rho\sigma}(x_{2})~B_{a_{3}\nu}(x_{3})
\nonumber\\
+ \partial^{\mu}_{3}T(C_{a_{1}}^{\rho(0)}(x_{1}), u_{a_{2}}^{(0)}(x_{2}), \tilde{u}_{a_{3}}^{(0)}(x_{3}))~
v_{a_{1}\rho}(x_{1})~D_{a_{2}}(x_{2})~B_{a_{3}}(x_{3})
\nonumber\\
- \partial_{\sigma}^{2}T(C_{a_{1}}^{\rho(0)}(x_{1}), \tilde{u}_{a_{2}}^{(0)}(x_{2}), u_{a_{3}}^{(0)}(x_{3}))~
v_{a_{1}\rho}(x_{1})~B_{a_{2}}^{\sigma}(x_{2})~C_{a_{3}}^{\mu}(x_{3})
\nonumber\\
+ {1\over 2}~T(E_{a_{1}}^{\rho\sigma(0)}(x_{1}), v_{a_{2}\lambda}^{(0)}(x_{2}), v_{a_{3}\nu}^{(0)}(x_{3}))~
F_{a_{1}\sigma\rho}(x_{1})~C_{a_{2}}^{\lambda}(x_{2})~C_{a_{3}}^{\nu\mu}(x_{3})
\nonumber\\
- {1\over 2}~T(E_{a_{1}}^{\rho\sigma(0)}(x_{1}), v_{a_{2}\lambda}^{(0)}(x_{2}), F_{a_{3}}^{\nu\mu(0)}(x_{3}))~
F_{a_{1}\sigma\rho}(x_{1})~C_{a_{2}}^{\lambda}(x_{2})~B_{a_{3}\nu}(x_{3})
\nonumber\\
+ {1\over 4}~
T(E_{a_{1}}^{\rho_{1}\sigma_{1}(0)}(x_{1}), F_{a_{2}\sigma_{2}\rho_{2}}^{(0)}(x_{2}), v_{a_{3}\nu}^{(0)}(x_{3}))~
F_{a_{1}\sigma_{1}\rho_{1}}(x_{1})~E_{a_{2}}^{\rho_{2}\sigma_{2}}(x_{2})~C_{a_{3}}^{\nu\mu}(x_{3})
\nonumber\\
- {1\over 4}~
T(E_{a_{1}}^{\rho_{1}\sigma_{1}(0)}(x_{1}), F_{a_{2}\sigma_{2}\rho_{2}}^{(0)}(x_{2}), F_{a_{3}}^{\nu\mu(0)}(x_{3}))~
F_{a_{1}\sigma_{1}\rho_{1}}(x_{1})~E_{a_{2}}^{\rho_{2}\sigma_{2}}(x_{2})~B_{a_{3}\nu}(x_{3})
\nonumber\\
+ T(D_{a_{1}}^{(0)}(x_{1}), v_{a_{2}\rho}^{(0)}(x_{2}), u_{a_{3}}^{(0)}(x_{3}))~
u_{a_{1}}(x_{1})~C_{a_{2}}^{\rho}(x_{2})~C_{a_{3}}^{\mu}(x_{3})
\nonumber\\
+ {1\over 2}~T(D_{a_{1}}^{(0)}(x_{1}), F_{a_{2}\sigma\rho}^{(0)}(x_{2}), u_{a_{3}}^{(0)}(x_{3}))~
u_{a_{1}\rho}(x_{1})~E_{a_{2}}^{\rho\sigma}(x_{2})~C_{a_{3}}^{\mu}(x_{3})
\nonumber\\
+ T(D_{a_{1}}^{(0)}(x_{1}), u_{a_{2}}^{(0)}(x_{2}), v_{a_{3}\nu}^{(0)}(x_{3}))~
u_{a_{1}}(x_{1})~D_{a_{2}}(x_{2})~C_{a_{3}}^{\nu\mu}(x_{3})
\nonumber\\
- T(D_{a_{1}}^{(0)}(x_{1}), u_{a_{2}}^{(0)}(x_{2}), F_{a_{3}}^{\nu\mu(0)}(x_{3}))~
u_{a_{1}}(x_{1})~D_{a_{2}}(x_{2})~B_{a_{3}\nu}(x_{3})
\nonumber\\
- \partial^{\mu}_{3}T(B_{a_{1}}^{\rho(0)}(x_{1}), v_{a_{2}\sigma}^{(0)}(x_{2}), \tilde{u}_{a_{3}}^{(0)}(x_{3}))~
\partial_{\rho}\tilde{u}_{a_{1}}(x_{1})~C_{a_{2}}^{\sigma}(x_{2})~B_{a_{3}}(x_{3})
\nonumber\\
- {1\over 2}~\partial^{\mu}_{3}
T(B_{a_{1}}^{\lambda(0)}(x_{1}), F_{a_{2}\sigma\rho}^{(0)}(x_{2}), \tilde{u}_{a_{3}}^{(0)}(x_{3}))~
\partial_{\lambda}\tilde{u}_{a_{1}}(x_{1})~E_{a_{2}}^{\rho\sigma}(x_{2})~B_{a_{3}}(x_{3})
\nonumber\\
+ \partial^{\sigma}_{2}T(B_{a_{1}}^{\rho(0)}(x_{1}), \tilde{u}_{a_{2}}^{(0)}(x_{2}), v_{a_{3}\nu}^{(0)}(x_{3}))~
\partial_{\lambda}\tilde{u}_{a_{1}}(x_{1})~B_{a_{2}}^{\sigma}(x_{2})~C_{a_{3}}^{\nu\mu}(x_{3})
\nonumber\\
- \partial^{\sigma}_{2}T(B_{a_{1}}^{\rho(0)}(x_{1}), \tilde{u}_{a_{2}}^{(0)}(x_{2}), F_{a_{3}}^{\nu\mu(0)}(x_{3}))~
\partial_{\lambda}\tilde{u}_{a_{1}}(x_{1})~B_{a_{2}}^{\sigma}(x_{2})~B_{a_{3}\nu}(x_{3})
\label{1,2,2mu}
\eea
\newpage
\bea
T(T^{\mu_{1}(2)}(x_{1}),T^{\mu_{2}(2)}(x_{2}),T^{(1)}(x_{3})) = 
\nonumber\\
T(v_{a_{1}\nu_{1}}^{(0)}(x_{1}), v_{a_{2}\nu_{2}}^{(0)}(x_{2}), C_{a_{3}}^{\rho(0)}(x_{3}))~
C_{a_{1}}^{\nu_{1}\mu_{1}}(x_{1})~C_{a_{2}}^{\nu_{2}\mu_{2}}(x_{2})~v_{a_{3}\rho}(x_{3})
\nonumber\\
+ {1\over 2}~T(v_{a_{1}\nu_{1}}^{(0)}(x_{1}), v_{a_{2}\nu_{2}}^{(0)}(x_{2}), E_{a_{3}}^{\rho\sigma(0)}(x_{3}))~
C_{a_{1}}^{\nu_{1}\mu_{1}}(x_{1})~C_{a_{2}}^{\nu_{2}\mu_{2}}(x_{2})~F_{a_{3}\sigma\rho}(x_{3})
\nonumber\\
+ T(F_{a_{1}}^{\nu_{1}\mu_{1}(0)}(x_{1}), F_{a_{2}}^{\nu_{2}\mu_{2}(0)}(x_{2}), C_{a_{3}}^{\rho(0)}(x_{3}))~
B_{a_{1}\nu_{1}}(x_{1})~B_{a_{2}\nu_{2}}(x_{2})~v_{a_{3}\rho}(x_{3})
\nonumber\\
+ {1\over 2}~T(F_{a_{1}}^{\nu_{1}\mu_{1}(0)}(x_{1}), F_{a_{2}}^{\nu_{2}\mu_{2}(0)}(x_{2}), E_{a_{3}}^{\rho\sigma(0)}(x_{3}))~
B_{a_{1}\nu_{1}}(x_{1})~B_{a_{2}\nu_{2}}(x_{2})~F_{a_{3}\sigma\rho}(x_{3})
\nonumber\\
+ T(u_{a_{1}}^{(0)}(x_{1}), v_{a_{2}\nu}^{(0)}(x_{2}), D_{a_{3}}^{(0)}(x_{3}))~
C_{a_{1}}^{\mu_{1}}(x_{1})~C_{a_{2}}^{\nu\mu_{2}}(x_{2})~u_{a_{3}}(x_{3}) 
- (x_{1}\leftrightarrow x_{2},\mu_{1} \leftrightarrow \mu_{2})
\nonumber\\
- T(u_{a_{1}}^{(0)}(x_{1}), F_{a_{2}}^{\nu\mu_{2}(0)}(x_{2}), D_{a_{3}}^{(0)}(x_{3}))~
C_{a_{1}}^{\mu_{1}}(x_{1})~B_{a_{2}\nu}(x_{2})~u_{a_{3}}(x_{3}) - (x_{1}\leftrightarrow x_{2},\mu_{1} \leftrightarrow \mu_{2})
\nonumber\\
- \partial^{\mu_{2}}_{2}~T(u_{a_{1}}^{(0)}(x_{1}), \tilde{u}_{a_{2}}^{(0)}(x_{2}), C_{a_{3}}^{\rho(0)}(x_{3}))~
C_{a_{1}}^{\mu_{1}}(x_{1})~B_{a_{2}}(x_{2})~v_{a_{3}\rho}(x_{3}) - (x_{1}\leftrightarrow x_{2},\mu_{1} \leftrightarrow \mu_{2})
\nonumber\\
- T(v_{a_{1}\nu}^{(0)}(x_{1}), F_{a_{2}}^{\lambda\mu_{2}(0)}(x_{2}), C_{a_{3}}^{\rho(0)}(x_{3}))~
C_{a_{1}}^{\nu\mu_{1}}(x_{1})~B_{a_{2}\lambda}(x_{2})~v_{a_{3}\rho}(x_{3}) 
- (x_{1}\leftrightarrow x_{2},\mu_{1} \leftrightarrow \mu_{2})
\nonumber\\
- {1\over 2}~T(v_{a_{1}\nu}^{(0)}(x_{1}), F_{a_{2}}^{\lambda\mu_{2}(0)}(x_{2}), E_{a_{3}}^{\rho\sigma(0)}(x_{3}))~
C_{a_{1}}^{\nu\mu_{1}}(x_{1})~B_{a_{2}\lambda}(x_{2})~F_{a_{3}\sigma\rho}(x_{3})
- (x_{1}\leftrightarrow x_{2},\mu_{1} \leftrightarrow \mu_{2})
\nonumber\\
+ \partial^{\mu_{2}}_{2}~T(v_{a_{1}\nu}^{(0)}(x_{1}), \tilde{u}_{a_{2}}^{(0)}(x_{2}), B_{a_{3}}^{\rho(0)}(x_{3}))~
C_{a_{1}}^{\nu\mu_{1}}(x_{1})~B_{a_{2}}(x_{2})~\partial_{\rho}\tilde{u}_{a_{3}}(x_{3})
- (x_{1}\leftrightarrow x_{2},\mu_{1} \leftrightarrow \mu_{2})
\nonumber\\
- \partial^{\mu_{2}}_{2}~T(F_{a_{1}}^{\nu\mu_{1}(0)}(x_{1}), \tilde{u}_{a_{2}}^{(0)}(x_{2}), B_{a_{3}}^{\rho(0)}(x_{3}))~
B_{a_{1}\nu}(x_{1})~B_{a_{2}}(x_{2})~\partial_{\rho}\tilde{u}_{a_{3}}(x_{3})
- (x_{1}\leftrightarrow x_{2},\mu_{1} \leftrightarrow \mu_{2})
\label{2mu,2nu,1}
\eea
\bea
T(T^{\mu_{1}(2)}(x_{1}),T^{\mu_{2}(1)}(x_{2}),T^{(2)}(x_{3})) = 
\nonumber\\
- \partial_{\rho}^{3}~T(u_{a_{1}}^{(0)}(x_{1}), C_{a_{2}}^{\mu_{2}(0)}(x_{2}), \tilde{u}_{a_{3}}^{(0)}(x_{3}))~
C_{a_{1}}^{\mu_{1}}(x_{1})~u_{a_{2}}(x_{2})~B_{a_{3}}^{\rho}(x_{3})
\nonumber\\
+ T(v_{a_{1}\nu}^{(0)}(x_{1}), C_{a_{2}}^{\mu_{2}(0)}(x_{2}), v_{a_{3}\rho}^{(0)}(x_{3}))~
C_{a_{1}}^{\nu\mu_{1}}(x_{1})~u_{a_{2}}(x_{2})~C_{a_{3}}^{\rho}(x_{3})
\nonumber\\
+ {1\over 2}~T(v_{a_{1}\nu}^{(0)}(x_{1}), C_{a_{2}}^{\mu_{2}(0)}(x_{2}), F_{a_{3}\sigma\rho}^{(0)}(x_{3}))~
C_{a_{1}}^{\nu\mu_{1}}(x_{1})~u_{a_{2}}(x_{2})~E_{a_{3}}^{\rho\sigma}(x_{3})
\nonumber\\
+ \partial_{\rho}^{3}~T(v_{a_{1}\lambda}^{(0)}(x_{1}), C_{a_{2}}^{\nu\mu_{2}(0)}(x_{2}), \tilde{u}_{a_{3}}(x_{3}))~
C_{a_{1}}^{\lambda\mu_{1}}(x_{1})~v_{a_{2}\nu}(x_{2})~B_{a_{3}}^{\rho}(x_{3})
\nonumber\\
+ \partial_{\sigma}^{3}~T(v_{a_{1}\nu}^{(0)}(x_{1}), B_{a_{2}\rho}^{(0)}(x_{2}), \tilde{u}_{a_{3}}^{(0)}(x_{3}))~
C_{a_{1}}^{\nu\mu_{1}}(x_{1})~F_{a_{2}}^{\mu_{2}\rho}(x_{2})~B_{a_{3}}^{\sigma}(x_{3}) 
\nonumber\\
- T(F_{a_{1}}^{\nu\mu_{1}(0)}(x_{1}), C_{a_{2}}^{\mu_{2}(0)}(x_{2}), v_{a_{3}\rho}^{(0)}(x_{3}))~
B_{a_{1}\nu}(x_{1})~u_{a_{2}}(x_{2})~C_{a_{3}}^{\rho}(x_{3}) 
\nonumber\\
- {1\over 2}~T(F_{a_{1}}^{\nu\mu_{1}(0)}(x_{1}), C_{a_{2}}^{\mu_{2}(0)}(x_{2}), F_{a_{3}\sigma\rho}^{(0)}(x_{3}))~
B_{a_{1}\nu}(x_{1})~u_{a_{2}}(x_{2})~E_{a_{3}}^{\rho\sigma}(x_{3})
\nonumber\\
- \partial_{\sigma}^{3}~T(F_{a_{1}}^{\nu\mu_{1}(0)}(x_{1}), C_{a_{2}}^{\rho\mu_{2}(0)}(x_{2}), \tilde{u}_{a_{3}}^{(0)}(x_{3}))~
B_{a_{1}\nu}(x_{1})~v_{a_{2}\rho}(x_{2})~B_{a_{3}}^{\sigma}(x_{3})
\nonumber\\
- \partial_{\sigma}^{3}~T(F_{a_{1}}^{\nu\mu_{1}(0)}(x_{1}), B_{a_{2}\rho}^{(0)}(x_{2}), \tilde{u}_{a_{3}}^{(0)}(x_{3}))~
B_{a_{1}\nu}(x_{1})~F_{a_{2}}^{\mu_{2}\rho}(x_{2})~B_{a_{3}}^{\sigma}(x_{3})
\nonumber\\
+ \partial^{\mu_{1}}_{1}~T(\tilde{u}_{a_{1}}^{(0)}(x_{1}), C_{a_{2}}^{\mu_{2}(0)}(x_{2}), u_{a_{3}}^{(0)}(x_{3}))~
B_{a_{1}}(x_{1})~u_{a_{2}}(x_{2})~D_{a_{3}}(x_{3})
\nonumber\\
- \partial^{\mu_{1}}_{1}~T(\tilde{u}_{a_{1}}^{(0)}(x_{1}), C_{a_{2}}^{\rho\mu_{2}(0)}(x_{2}), v_{a_{3}\sigma}^{(0)}(x_{3}))~
B_{a_{1}}(x_{1})~v_{a_{2}\rho}(x_{2})~C_{a_{3}}^{\sigma}(x_{3})
\nonumber\\
- {1\over 2}~\partial^{\mu_{1}}_{1}~
T(\tilde{u}_{a_{1}}^{(0)}(x_{1}), C_{a_{2}}^{\nu\mu_{2}(0)}(x_{2}), F_{a_{3}\sigma\rho}^{(0)}(x_{3}))~
B_{a_{1}}(x_{1})~v_{a_{2}\nu}(x_{2})~E_{a_{3}}^{\rho\sigma}(x_{3})
\nonumber\\
- \partial^{\mu_{1}}_{1}~T(\tilde{u}_{a_{1}}^{(0)}(x_{1}), B_{a_{2}\nu}^{(0)}(x_{2}), v_{a_{3}\sigma}^{(0)}(x_{3}))~
B_{a_{1}}(x_{1})~F_{a_{2}}^{\mu_{2}\nu}(x_{2})~C_{a_{3}}^{\sigma}(x_{3})
\nonumber\\
- {1\over 2}~\partial^{\mu_{1}}_{1}~
T(\tilde{u}_{a_{1}}^{(0)}(x_{1}), B_{a_{2}\nu}^{(0)}(x_{2}), F_{a_{3}\sigma\rho}^{(0)}(x_{3}))~
B_{a_{1}}(x_{1})~F_{a_{2}}^{\mu_{2}\nu}(x_{2})~E_{a_{3}}^{\rho\sigma}(x_{3})
\nonumber\\
- \partial^{\mu_{1}}_{1}\partial_{\nu}^{3}~T(\tilde{u}_{a_{1}}^{(0)}(x_{1}), B_{a_{2}}^{(0)}(x_{2}), \tilde{u}_{a_{3}}^{(0)}(x_{3}))~
B_{a_{1}}(x_{1})~\partial^{\mu_{2}}\tilde{u}_{a_{2}}(x_{2})~B_{a_{3}}^{\nu}(x_{3})
\label{2mu,1nu,2}
\eea
\bea
T(T^{(2)}(x_{1}),T^{(2)}(x_{2}),T^{\mu\nu(1)}(x_{3})) = 
\nonumber\\
- \partial_{\rho_{1}}^{1}\partial_{\rho_{2}}^{2}~
T(\tilde{u}_{a_{1}}^{(0)}(x_{1}), \tilde{u}_{a_{2}}^{(0)}(x_{2}), B_{a_{3}}^{(0)}(x_{3}))~
B_{a_{1}}^{\rho_{1}}(x_{1})~B_{a_{2}}^{\rho_{2}}(x_{2})~F_{a_{3}}^{\mu\nu}(x_{3})
\nonumber\\
- \partial_{\sigma}^{2}~T(v_{a_{1}\sigma}^{(0)}(x_{1}), \tilde{u}_{a_{2}}^{(0)}(x_{2}), C_{a_{3}}^{\mu\nu(0)}(x_{3}))~
C_{a_{1}}^{\rho}(x_{1})~B_{a_{2}}^{\sigma}(x_{2})~u_{a_{3}}(x_{3}) + (x_{1}\leftrightarrow x_{2})
\nonumber\\
- {1\over 2}~\partial_{\lambda}^{2}~T(F_{a_{1}\sigma\rho}^{(0)}(x_{1}), \tilde{u}_{a_{2}}^{(0)}(x_{2}), C_{a_{3}}^{\mu\nu(0)}(x_{3}))~
E_{a_{1}}^{\rho\sigma}(x_{1})~B_{a_{2}}^{\lambda}(x_{2})~u_{a_{3}}(x_{3}) + (x_{1}\leftrightarrow x_{2})
\label{2,2,1munu}
\eea
\bea
T(T^{(1)}(x_{1}),T^{(2)}(x_{2}),T^{\mu\nu(2)}(x_{3})) = 
\nonumber\\
T(C_{a_{1}}^{\rho(0)}(x_{1}), v_{a_{2}\sigma}^{(0)}(x_{2}), F_{a_{3}}^{\mu\nu(0)}(x_{3}))~
v_{a_{1}\rho}(x_{1})~C_{a_{2}}^{\sigma}(x_{2})~B_{a_{3}}(x_{3})
\nonumber\\
+ {1\over 2}~T(C_{a_{1}}^{\lambda(0)}(x_{1}), F_{a_{2}\sigma\rho}^{(0)}(x_{2}), F_{a_{3}}^{\mu\nu(0)}(x_{3}))~
v_{a_{1}\lambda}(x_{1})~E_{a_{2}}^{\rho\sigma}(x_{2})~B_{a_{3}}(x_{3})
\nonumber\\
+ \partial_{\sigma}^{2}~T(C_{a_{1}}^{\rho(0)}(x_{1}), \tilde{u}_{a_{2}}^{(0)}(x_{2}), u_{a_{3}}^{(0)}(x_{3}))~
v_{a_{1}\rho}(x_{1})~B_{a_{2}}^{\sigma}(x_{2})~C_{a_{3}}^{\mu\nu}(x_{3})
\nonumber\\
+ {1\over 2}~T(E_{a_{1}}^{\rho\sigma(0)}(x_{1}), v_{a_{2}\lambda}^{(0)}(x_{2}), F_{a_{3}}^{\mu\nu(0)}(x_{3}))~
F_{a_{1}\sigma\rho}(x_{1})~C_{a_{2}}^{\lambda}(x_{2})~B_{a_{3}}(x_{3})
\nonumber\\
+ {1\over 4}~T(E_{a_{1}}^{\rho_{1}\sigma_{1}(0)}(x_{1}), F_{a_{2}\sigma_{2}\rho_{2}}^{(0)}(x_{2}), F_{a_{3}}^{\mu\nu(0)}(x_{3}))~
F_{a_{1}\sigma_{1}\rho_{1}}(x_{1})~E_{a_{2}}^{\rho_{2}\sigma_{2}}(x_{2})~B_{a_{3}}(x_{3})
\nonumber\\
- T(D_{a_{1}}^{(0)}(x_{1}), v_{a_{2}\rho}^{(0)}(x_{2}), u_{a_{3}}^{(0)}(x_{3}))~
u_{a_{1}}(x_{1})~C_{a_{2}}^{\rho}(x_{2})~C_{a_{3}}^{\mu\nu}(x_{3})
\nonumber\\
- {1\over 2}~T(D_{a_{1}}^{(0)}(x_{1}), F_{a_{2}\sigma\rho}^{(0)}(x_{2}), u_{a_{3}}^{(0)}(x_{3}))~
u_{a_{1}}(x_{1})~E_{a_{2}}^{\rho\sigma}(x_{2})~C_{a_{3}}^{\mu\nu}(x_{3})
\nonumber\\
+ T(D_{a_{1}}^{(0)}(x_{1}), u_{a_{2}}^{(0)}(x_{2}), F_{a_{3}}^{\mu\nu(0)}(x_{3}))~
u_{a_{1}}(x_{1})~D_{a_{2}}(x_{2})~B_{a_{3}}(x_{3})
\nonumber\\
+ \partial_{\sigma}^{2}~T(B_{a_{1}}^{\rho(0)}(x_{1}), \tilde{u}_{a_{2}}^{(0)}(x_{2}), F_{a_{3}}^{\mu\nu(0)}(x_{3}))~
\partial_{\rho}\tilde{u}_{a_{1}}(x_{1})~B_{a_{2}}^{\sigma}(x_{2})~B_{a_{3}}(x_{3})
\label{1,2,2munu}
\eea
\newpage
\bea
T(T^{\mu_{1}(2)}(x_{1}),T^{\mu_{2}(2)}(x_{2}),T^{\mu_{3}(1)}(x_{3})) = 
\nonumber\\
T(v_{a_{1}\nu_{1}}^{(0)}(x_{1}), v_{a_{2}\nu_{2}}^{(0)}(x_{2}), C_{a_{3}}^{\mu_{3}(0)}(x_{3}))~
C_{a_{1}}^{\nu_{1}\mu_{1}}(x_{1})~C_{a_{2}}^{\nu_{2}\mu_{2}}(x_{2})~u_{a_{3}}(x_{3})
\nonumber\\
+ T(F_{a_{1}}^{\nu_{1}\mu_{1}(0)}(x_{1}), F_{a_{2}}^{\nu_{2}\mu_{2}(0)}(x_{2}), C_{a_{3}}^{\mu_{3}(0)}(x_{3}))~
B_{a_{1}\nu_{1}}(x_{1})~B_{a_{2}\nu_{2}}(x_{2})~u_{a_{3}}(x_{3})
\nonumber\\
- \partial^{\mu_{1}}_{1}~\partial^{\mu_{2}}_{2}~
T(\tilde{u}_{a_{1}}^{(0)}(x_{1}), \tilde{u}_{a_{2}}^{(0)}(x_{2}), B_{a_{3}}^{(0)}(x_{3}))~
B_{a_{1}}(x_{1})~B_{a_{2}}(x_{2})~\partial^{\mu_{3}}\tilde{u}_{a_{3}}(x_{3})
\nonumber\\
 - \partial^{\mu_{2}}_{2}~T(u_{a_{1}}^{(0)}(x_{1}), \tilde{u}_{a_{2}}^{(0)}(x_{2}), C_{a_{3}}^{\mu_{3}(0)}(x_{3}))~
C_{a_{1}}^{\mu_{1}}(x_{1})~B_{a_{2}}(x_{2})~u_{a_{3}}(x_{3}) 
- (x_{1}\leftrightarrow x_{2},\mu_{1} \leftrightarrow \mu_{2})
\nonumber\\
- T(v_{a_{1}\nu_{1}}^{(0)}(x_{1}), F_{a_{2}}^{\nu_{2}\mu_{2}(0)}(x_{2}), C_{a_{3}}^{\mu_{3}(0)}(x_{3}))~
C_{a_{1}}^{\nu_{1}\mu_{1}}(x_{1})~B_{a_{2}\nu_{2}}(x_{2})~u_{a_{3}}(x_{3}) 
- (x_{1}\leftrightarrow x_{2},\mu_{1} \leftrightarrow \mu_{2})
\nonumber\\
- \partial^{\mu_{2}}_{2}~T(v_{a_{1}\nu_{1}}^{(0)}(x_{1}), \tilde{u}_{a_{2}}^{(0)}(x_{2}), C_{a_{3}}^{\nu_{3}\mu_{3}(0)}(x_{3}))~
C_{a_{1}}^{\nu_{1}\mu_{1}}(x_{1})~B_{a_{2}}(x_{2})~v_{a_{3}\nu_{3}}(x_{3}) 
- (x_{1}\leftrightarrow x_{2},\mu_{1} \leftrightarrow \mu_{2})
\nonumber\\
- \partial^{\mu_{2}}_{2}~T(v_{a_{1}\nu_{1}}^{(0)}(x_{1}), \tilde{u}_{a_{2}}^{(0)}(x_{2}), B_{a_{3}\nu_{3}}^{(0)}(x_{3}))~
C_{a_{1}}^{\nu_{1}\mu_{1}}(x_{1})~B_{a_{2}}(x_{2})~F_{a_{3}}^{\mu_{3}\nu_{3}}(x_{3}) 
- (x_{1}\leftrightarrow x_{2},\mu_{1} \leftrightarrow \mu_{2})
\nonumber\\
+ \partial^{\mu_{2}}_{2}~
T(F_{a_{1}}^{\nu_{1}\mu_{1}(0)}(x_{1}), \tilde{u}_{a_{2}}^{(0)}(x_{2}), C_{a_{3}}^{\nu_{3}\mu_{3}(0)}(x_{3}))~
B_{a_{1}\nu_{1}}(x_{1})~B_{a_{2}}(x_{2})~v_{a_{3}\nu_{3}}(x_{3}) 
\nonumber\\
- (x_{1}\leftrightarrow x_{2},\mu_{1} \leftrightarrow \mu_{2})
\nonumber\\
+ \partial^{\mu_{2}}_{2}~T(F_{a_{1}}^{\nu_{1}\mu_{1}(0)}(x_{1}), \tilde{u}_{a_{2}}^{(0)}(x_{2}), B_{a_{3}\nu_{3}}^{(0)}(x_{3}))~
B_{a_{1}\nu_{1}}(x_{1})~B_{a_{2}}(x_{2})~F_{a_{3}}^{\mu_{3}\nu_{3}}(x_{3}) 
\nonumber\\
- (x_{1}\leftrightarrow x_{2},\mu_{1} \leftrightarrow \mu_{2})
\label{2mu,2nu,1rho}
\eea
\bea
T(T^{\mu\nu(2)}(x_{1}),T^{\rho(2)}(x_{2}),T^{(1)}(x_{3})) = 
\nonumber\\
- T(F_{a_{1}}^{\mu\nu(0)}(x_{1}), u_{a_{2}}^{(0)}(x_{2}), D_{a_{3}}^{(0)}(x_{3}))~
B_{a_{1}}(x_{1})~C_{a_{2}}^{\rho}(x_{2})~u_{a_{3}}(x_{3})
\nonumber\\
+ T(F_{a_{1}}^{\mu\nu(0)}(x_{1}), v_{a_{2}\sigma}^{(0)}(x_{2}), C_{a_{3}}^{\lambda(0)}(x_{3}))~
B_{a_{1}}(x_{1})~C_{a_{2}}^{\sigma\rho}(x_{2})~v_{a_{3}\lambda}(x_{3})
\nonumber\\
+ {1\over 2}~T(F_{a_{1}}^{\mu\nu(0)}(x_{1}), v_{a_{2}\sigma}^{(0)}(x_{2}), E_{a_{3}}^{\alpha\beta(0)}(x_{3}))~
B_{a_{1}}(x_{1})~C_{a_{2}}^{\sigma\rho}(x_{2})~F_{a_{3}\beta\alpha}(x_{3})
\nonumber\\
 - T(F_{a_{1}}^{\mu\nu(0)}(x_{1}), F_{a_{2}}^{\sigma\rho(0)}(x_{2}), C_{a_{3}}^{\lambda(0)}(x_{3}))~
B_{a_{1}}(x_{1})~B_{a_{2}\sigma}(x_{2})~v_{a_{3}\lambda}(x_{3}) 
\nonumber\\
- {1\over 2}~T(F_{a_{1}}^{\mu\nu(0)}(x_{1}), F_{a_{2}}^{\sigma\rho(0)}(x_{2}), E_{a_{3}}^{\alpha\beta(0)}(x_{3}))~
B_{a_{1}}(x_{1})~B_{a_{2}\sigma}(x_{2})~F_{a_{3}\beta\alpha}(x_{3})
\nonumber\\
+ \partial^{\rho}_{2}~T(F_{a_{1}}^{\mu\nu(0)}(x_{1}), \tilde{u}_{a_{2}}^{(0)}(x_{2}), B_{a_{3}}^{\sigma(0)}(x_{3}))~
B_{a_{1}}(x_{1})~B_{a_{2}}(x_{2})~\partial_{\sigma}\tilde{u}_{a_{3}}(x_{3}) 
\nonumber\\
+ T(u_{a_{1}}^{(0)}(x_{1}), v_{a_{2}\sigma}^{(0)}(x_{2}), D_{a_{3}}^{(0)}(x_{3}))~
C_{a_{1}}^{\mu\nu}(x_{1})~C_{a_{2}}^{\sigma\rho}(x_{2})~u_{a_{3}}(x_{3}) 
\nonumber\\
- T(u_{a_{1}}^{(0)}(x_{1}), F_{a_{2}}^{\sigma\rho(0)}(x_{2}), D_{a_{3}}^{(0)}(x_{3}))~
C_{a_{1}}^{\mu\nu}(x_{1})~B_{a_{2}\sigma}(x_{2})~u_{a_{3}}(x_{3}) 
\nonumber\\
- \partial^{\rho}_{2}~T(u_{a_{1}}^{(0)}(x_{1}), \tilde{u}_{a_{2}}^{(0)}(x_{2}), C_{a_{3}}^{\sigma(0)}(x_{3}))~
C_{a_{1}}^{\mu\nu}(x_{1})~B_{a_{2}}(x_{2})~v_{a_{3}\sigma}(x_{3})
\label{2munu,2rho,1}
\eea
\bea
T(T^{\mu\nu(2)}(x_{1}),T^{\rho(1)}(x_{2}),T^{(2)}(x_{3})) = 
\nonumber\\
T(F_{a_{1}}^{\mu\nu(0)}(x_{1}), C_{a_{2}}^{\rho(0)}(x_{2}), v_{a_{3}\sigma}^{(0)}(x_{3}))~
B_{a_{1}}(x_{1})~u_{a_{2}}(x_{2})~C_{a_{3}}^{\sigma}(x_{3})
\nonumber\\
+ {1\over 2}~T(F_{a_{1}}^{\mu\nu(0)}(x_{1}), C_{a_{2}}^{\rho(0)}(x_{2}), F_{a_{3}\beta\alpha}^{(0)}(x_{3}))~
B_{a_{1}}(x_{1})~u_{a_{2}}(x_{2})~E_{a_{3}}^{\alpha\beta}(x_{3})
\nonumber\\
+ \partial^{\lambda}_{3}~T(F_{a_{1}}^{\mu\nu(0)}(x_{1}), C_{a_{2}}^{\sigma\rho(0)}(x_{2}), \tilde{u}_{a_{3}}^{(0)}(x_{3}))~
B_{a_{1}}(x_{1})~v_{a_{2}\sigma}(x_{2})~B_{a_{3}\lambda}(x_{3})
\nonumber\\
+  \partial^{\lambda}_{3}~T(F_{a_{1}}^{\mu\nu(0)}(x_{1}), B_{a_{2}\sigma}^{(0)}(x_{2}), \tilde{u}_{a_{3}}^{(0)}(x_{3}))~
B_{a_{1}}(x_{1})~F_{a_{2}}^{\rho\sigma}(x_{2})~B_{a_{3}\lambda}(x_{3}) 
\nonumber\\
- \partial^{\sigma}_{3}~T(u_{a_{1}}^{(0)}(x_{1}), C_{a_{2}}^{\rho(0)}(x_{2}), \tilde{u}_{a_{3}}^{(0)}(x_{3}))~
C_{a_{1}}^{\mu\nu}(x_{1})~u_{a_{2}}(x_{2})~B_{a_{3}\sigma}(x_{3})
\label{2munu,1rho,2}
\eea
\newpage
\bea
T(T^{\mu\nu(1)}(x_{1}),T^{\rho(2)}(x_{2}),T^{(2)}(x_{3})) = 
\nonumber\\
\partial^{\lambda}_{3}~T(C_{a_{1}}^{\mu\nu(0)}(x_{1}), v_{a_{2}\sigma}^{(0)}(x_{2}), \tilde{u}_{a_{3}}^{(0)}(x_{3}))~
u_{a_{1}}(x_{1})~C_{a_{2}}^{\sigma\rho}(x_{2})~B_{a_{3}}^{\lambda}(x_{3})
\nonumber\\
+ \partial^{\lambda}_{3}~T(C_{a_{1}}^{\mu\nu(0)}(x_{1}), F_{a_{2}}^{\sigma\rho(0)}(x_{2}), \tilde{u}_{a_{3}}^{(0)}(x_{3}))~
u_{a_{1}}(x_{1})~B_{a_{2}\sigma}(x_{2})~B_{a_{3}}^{\lambda}(x_{3})
\nonumber\\
+ \partial^{\rho}_{2}~T(C_{a_{1}}^{\mu\nu(0)}(x_{1}), \tilde{u}_{a_{2}}^{(0)}(x_{2}), v_{a_{3}\sigma}^{(0)}(x_{3}))~
u_{a_{1}}(x_{1})~B_{a_{2}}(x_{2})~C_{a_{3}}^{\sigma}(x_{3})
\nonumber\\
+ {1\over 2}~\partial^{\rho}_{2}~T(C_{a_{1}}^{\mu\nu(0)}(x_{1}), \tilde{u}_{a_{2}}^{(0)}(x_{2}), F_{a_{3}\beta\alpha}^{(0)}(x_{3}))~
u_{a_{1}}(x_{1})~B_{a_{2}}(x_{2})~E_{a_{3}}^{\alpha\beta}(x_{3}) 
\nonumber\\
- \partial^{\rho}_{2}\partial^{\sigma}_{3}~T(B_{a_{1}}^{(0)}(x_{1}), \tilde{u}_{a_{2}}^{(0)}(x_{2}), \tilde{u}_{a_{3}}^{(0)}(x_{3}))~
F_{a_{1}}^{\mu\nu}(x_{1})~B_{a_{2}}(x_{2})~B_{a_{3}\sigma}(x_{3})
\label{1munu,2rho,2}
\eea

All the coefficients from the 11 equations from above are in fact chronological products of the type
$
T(\xi_{1}^{(0)}(x_{1}), \xi_{2}^{(0)}(x_{2}), (\eta_{1}\eta_{2})^{(0)}(x_{3})).
$
In the simplest case when all variables are even we have
\bea
T(\xi_{1}^{(0)}(x_{1}), \xi_{2}^{(0)}(x_{2}), (\eta_{1}\eta_{2})^{(0)}(x_{3}))
\nonumber\\
= D^{F}(\xi_{1}(x_{1}), \eta_{1}(x_{3}))~D^{F}(\xi_{2}(x_{2}), \eta_{2}(x_{3}))
+ D^{F}(\xi_{1}(x_{1}), \eta_{2}(x_{3}))~D^{F}(\xi_{2}(x_{2}), \eta_{1}(x_{3}))
\eea
and in the general case one has to introduce apropiate signs taking care of the Grassmann parities.

One can see from this formula that there will be two contributions: a {\it canonical splitting} contributions given by 
(\ref{canonical}) i.e. we can pull out the derivatives from the chronological product, and a {\it finite renormalization} contribution
due to (\ref{vv}) and from this relation
\be
N(F_{a}^{\mu\nu},F_{b}^{\rho\sigma}) = i~(\eta^{\mu\rho}~\eta^{\nu\sigma} - \eta^{\nu\rho}~\eta^{\mu\sigma})~\delta_{ab}.
\label{N-FF}
\ee

From these formulas we get
\bea
T(F_{a_{1}}^{\mu\nu(0)}(x_{1}), v_{a_{2}}^{\rho(0)}(x_{2}), C_{a_{3}}^{\sigma(0)}(x_{3}))^{\rm ren}
= \partial^{\mu}_{1}T(v_{a_{1}}^{\nu(0)}(x_{1}), v_{a_{2}}^{\rho(0)}(x_{2}), C_{a_{3}}^{\sigma(0)}(x_{3})) 
- (\mu \leftrightarrow \nu)
\nonumber\\
+ f_{a_{1}a_{2}a_{3}}~\eta^{\mu\rho}~\eta^{\nu\sigma} ~\delta(x_{1} - x_{3})
D^{F}_{0}(x_{2} - x_{3}) - (\mu \leftrightarrow \nu)
\eea
\bea
T(F_{a_{1}}^{\mu\nu(0)}(x_{1}), F_{a_{2}}^{\rho\sigma(0)}(x_{2}), C_{a_{3}}^{\lambda(0)}(x_{3}))^{\rm ren}
\nonumber\\
= [ \partial^{\mu}_{1}\partial^{\rho}_{2}T(v_{a_{1}}^{\nu(0)}(x_{1}), v_{a_{2}}^{\sigma(0)}(x_{2}), C_{a_{3}}^{\lambda(0)}(x_{3})) 
- (\mu \leftrightarrow \nu) ] - (\rho \leftrightarrow \sigma)
\nonumber\\
- f_{a_{1}a_{2}a_{3}}~[ (\eta^{\nu\rho}~\eta^{\sigma\lambda}\partial^{\mu}D^{F}_{0}(x_{1} - x_{3})~\delta(x_{2} - x_{3}) 
\nonumber\\
- \eta^{\mu\sigma}~\eta^{\nu\lambda}\delta(x_{1} - x_{3})~\partial^{\rho}D^{F}_{0}(x_{2} - x_{3}) 
- (\mu \leftrightarrow \nu) ] - (\rho \leftrightarrow \sigma)
\eea
and
\bea
T(\tilde{u}_{a_{1}}^{(0)}(x_{1}), F_{a_{2}}^{\mu\nu(0)}(x_{2}), C_{a_{3}}^{\rho\sigma(0)}(x_{3}))^{\rm ren}
\nonumber\\
=  \partial^{\mu}_{2}T(\tilde{u}_{a_{1}}^{(0)}(x_{1}), v_{a_{2}}^{\nu(0)}(x_{2}), C_{a_{3}}^{\rho\sigma(0)}(x_{3})) 
- (\mu \leftrightarrow \nu) 
\nonumber\\
+ f_{a_{1}a_{2}a_{3}}~\eta^{\mu\rho}~\eta^{\nu\sigma} D^{F}_{0}(x_{1} - x_{3})~\delta(x_{2} - x_{3}) 
- (\mu \leftrightarrow \nu) 
\eea
where one can see the two contributions. Using these three relations we can obtain the finite renormalization contributions to the 
relations (\ref{2,2,1}) - (\ref{1munu,2rho,2}).
\begin{thm}
Following the finite renormalization 
\bea
D(v_{a\mu,\nu}(x),v_{b\rho,\sigma}(y))^{f} = {i \over 2}~\eta_{\mu\rho}~\eta_{\nu\sigma}~\delta_{ab}~\delta(x - y)
\Rightarrow
\nonumber\\
D(F_{a}^{\mu\nu}(x),F_{b}^{\rho\sigma}(y))^{f} = i~(\eta_{\mu\rho}~\eta_{\nu\sigma} - \eta_{\nu\rho}~\eta_{\mu\sigma})~\delta_{ab}~
\delta(x - y)
\eea
we have the following finite renormalizations of the expressions (\ref{2,2,1}) - (\ref{1munu,2rho,2}):
\bea
T(T^{(2)}(x_{1}),T^{(2)}(x_{2}),T^{(1)}(x_{3}))^{f} = 
\nonumber\\
f_{a_{1}a_{2}a_{3}}~[ \delta(x_{1} - x_{3})~D^{F}_{0}(x_{2} - x_{3})~
(E_{a_{1}}^{\rho\sigma}v_{a_{3}\rho})(x_{1})~C_{a_{2}\sigma}(x_{2}) 
\nonumber\\
- \delta(x_{1} - x_{3})~\partial_{\mu}D^{F}_{0}(x_{2} - x_{3})~
(E_{a_{1}\nu\rho}v_{a_{3}^{\rho}})(x_{1})~E_{a_{2}}^{\nu\mu}(x_{2}) ]
+ (x_{1} \leftrightarrow x_{2})
\eea
\bea
T(T^{(2)}(x_{1}),T^{(2)}(x_{2}),T^{\mu(1)}(x_{3}))^{f} = 
\nonumber\\
f_{a_{1}a_{2}a_{3}}~\{ - \delta(x_{1} - x_{3})~D^{F}_{0}(x_{2} - x_{3})~
(E_{a_{1}}^{\nu\mu}u_{a_{3}})(x_{1})~C_{a_{2}\nu}(x_{2}) 
\nonumber\\
+ \delta(x_{1} - x_{3})~\partial_{\nu}D^{F}_{0}(x_{2} - x_{3})~
[ (E_{a_{2}}^{\rho\mu}v_{a_{3}\rho})(x_{1})~B_{a_{1}}^{\nu}(x_{2})
+ ({E_{a_{1}}^{\mu}}_{\rho}u_{a_{3}})(x_{1})~E_{a_{2}}^{\rho\nu}(x_{2})] \}
\nonumber\\
+ (x_{1} \leftrightarrow x_{2})
\eea
\bea
T(T^{(1)}(x_{1}),T^{(2)}(x_{2}),T^{\mu(2)}(x_{3}))^{f} = 
\nonumber\\
f_{a_{1}a_{2}a_{3}}~\{ \delta(x_{3} - x_{1})~D^{F}_{0}(x_{2} - x_{1})~
[ (v_{a_{1}}^{\mu}B_{a_{3}\nu})(x_{3})~C_{a_{2}}^{\nu}(x_{2}) 
- (v_{a_{1}}^{\nu}B_{a_{3}\nu})(x_{3})~C_{a_{2}}^{\mu}(x_{2}) ]
\nonumber\\
- \delta(x_{2} - x_{1})~D^{F}_{0}(x_{3} - x_{1})~
(v_{a_{1}}^{\lambda}E_{a_{2}\nu\lambda})(x_{2})~C_{a_{2}}^{\nu\mu}(x_{3})
\nonumber\\
- \delta(x_{3} - x_{1})~\partial_{\nu}D^{F}_{0}(x_{2} - x_{1})~
[ (v_{a_{1}}^{\rho}B_{a_{3}\rho})(x_{3})~E_{a_{2}}^{\mu\nu}(x_{2}) 
- (v_{a_{1}}^{\mu}B_{a_{3}\rho})(x_{3})~E_{a_{2}}^{\rho\nu}(x_{2}) ]
\nonumber\\
- \delta(x_{2} - x_{1})~\partial_{\nu}D^{F}_{0}(x_{3} - x_{1})~
(v_{a_{1}\rho}E_{a_{2}}^{\rho\mu})(x_{2})~B_{a_{3}}^{\nu}(x_{3})
\nonumber\\
+ \delta(x_{2} - x_{1})~\partial^{\mu}D^{F}_{0}(x_{3} - x_{1})~
(v_{a_{1}\rho}E_{a_{2}}^{\rho\nu})(x_{2})~B_{a_{3}\nu}(x_{3})\}
\eea
\bea
T(T^{\mu_{1}(2)}(x_{1}),T^{\mu_{2}(2)}(x_{2}),T^{(1)}(x_{3}))^{f} = 
\nonumber\\
- f_{a_{1}a_{2}a_{3}}~\{ \delta(x_{1} - x_{3})~D^{F}_{0}(x_{2} - x_{3})~
[ (B_{a_{2}}^{\rho}v_{a_{3}\rho})(x_{1})~C_{a_{1}}^{\mu_{1}\mu_{2}}(x_{2}) 
- (B_{a_{1}\nu}v_{a_{3}}^{\mu_{1}})(x_{1})~C_{a_{1}}^{\nu\mu_{2}}(x_{2}) ]
\nonumber\\
- \delta(x_{1} - x_{3})~\partial^{\mu_{2}}D^{F}_{0}(x_{2} - x_{3})~
[ (B_{a_{1}\rho}v_{a_{3}}^{\mu_{1}})(x_{1})~B_{a_{2}}^{\rho}(x_{2})
- (B_{a_{1}}^{\rho}v_{a_{3}\rho})(x_{1})~B_{a_{2}}^{\mu_{1}}(x_{2}) ]
\nonumber\\
+ \delta(x_{1} - x_{3})~\partial_{\lambda}D^{F}_{0}(x_{2} - x_{3})~
(\eta^{\mu_{1}\mu_{2}}~B_{a_{1}}^{\rho}v_{a_{3}\rho} 
- B_{a_{1}}^{\mu_{2}}v_{a_{3}}^{\mu_{1}})(x_{1})~B_{a_{2}}^{\lambda}(x_{2}) \}
\nonumber\\
- (x_{1} \leftrightarrow x_{2}, \mu_{1} \leftrightarrow \mu_{2})
\eea
\bea
T(T^{\mu_{1}(2)}(x_{1}),T^{\mu_{2}(1)}(x_{2}),T^{(2)}(x_{3}))^{f} = 
\nonumber\\
f_{a_{1}a_{2}a_{3}}~\{ \delta(x_{3} - x_{2})~D^{F}_{0}(x_{1} - x_{2})~
C_{a_{1}}^{\nu\mu_{1}}(x_{1})~(u_{a_{2}}{E_{a_{3}}^{\mu_{2}}}_{\nu})(x_{3}) 
\nonumber\\
- \delta(x_{1} - x_{2})~D^{F}_{0}(x_{3} - x_{2})~
[ (B_{a_{1}}^{\mu_{2}}u_{a_{2}})(x_{1})~C_{a_{3}}^{\mu_{1}}(x_{3})
- \eta^{\mu_{1}\mu_{2}}~(B_{a_{1}\nu}u_{a_{2}})(x_{1})~C_{a_{3}}^{\nu}(x_{3}) ]
\nonumber\\
- \delta(x_{3} - x_{2})~\partial_{\nu}D^{F}_{0}(x_{1} - x_{2})~
B_{a_{1}}^{\nu}(x_{1})~(u_{a_{2}}E_{a_{3}}^{\mu_{1}\mu_{2}})(x_{3}) 
\nonumber\\
+ \delta(x_{3} - x_{2})~\partial^{\mu_{1}}D^{F}_{0}(x_{1} - x_{2})~
[ B_{a_{1}\nu}(x_{1})~(u_{a_{2}}E_{a_{3}}^{\mu_{2}\nu})(x_{3}) 
+ B_{a_{1}}(x_{1})~(v_{a_{2}\nu}E_{a_{3}}^{\mu_{2}\nu})(x_{3}) ] 
\nonumber\\
- \delta(x_{1} - x_{2})~\partial_{\nu}D^{F}_{0}(x_{3} - x_{2})~
[ (B_{a_{1}}^{\mu_{2}}u_{a_{2}})(x_{1})~E_{a_{3}}^{\mu_{1}\nu}(x_{3}) 
- \eta^{\mu_{1}\mu_{2}}~(B_{a_{1}\rho}u_{a_{2}})(x_{1})~E_{a_{3}}^{\rho\nu}(x_{3})
\nonumber\\
- (\eta^{\mu_{1}\mu_{2}}~B_{a_{1}\rho}v_{a_{2}}^{\rho} 
- B_{a_{1}}^{\mu_{2}}v_{a_{2}}^{\mu_{1}})(x_{1})~B_{a_{3}}^{\nu}(x_{3}) ] \}
\eea
\bea
T(T^{(2)}(x_{1}),T^{(2)}(x_{2}),T^{\mu\nu(1)}(x_{3}))^{f} = 
\nonumber\\
f_{a_{1}a_{2}a_{3}}~[ \delta(x_{1} - x_{3})~\partial_{\rho}D^{F}_{0}(x_{2} - x_{3})~
(E_{a_{1}}^{\mu\nu}u_{a_{3}})(x_{1})~B_{a_{2}}^{\rho}(x_{2}) 
+ (x_{1} \leftrightarrow x_{2})
\eea
\bea
T(T^{(1)}(x_{1}),T^{(2)}(x_{2}),T^{\mu\nu(2)}(x_{3}))^{f} = 
\nonumber\\
f_{a_{1}a_{2}a_{3}}~[ - \delta(x_{3} - x_{1})~D^{F}_{0}(x_{2} - x_{1})~
(v_{a_{1}}^{\nu}B_{a_{3}})(x_{3})~C_{a_{2}}^{\mu}(x_{2}) 
\nonumber\\
- \delta(x_{3} - x_{1})~\partial_{\rho}D^{F}_{0}(x_{2} - x_{1})~
(v_{a_{1}}^{\nu}B_{a_{3}})(x_{3})~E_{a_{2}}^{\mu\rho}(x_{2})
\nonumber\\
- \delta(x_{2} - x_{1})~\partial^{\nu}D^{F}_{0}(x_{3} - x_{1})~
(v_{a_{1}\rho}E_{a_{2}}^{\rho\mu})(x_{2})~B_{a_{3}}(x_{3}) ]
- (\mu \leftrightarrow \nu)
\eea
\bea
T(T^{\mu_{1}(2)}(x_{1}),T^{\mu_{2}(2)}(x_{2}),T^{\mu_{3}(1)}(x_{3}))^{f} = 
\nonumber\\
- f_{a_{1}a_{2}a_{3}}~\{ \delta(x_{1} - x_{3})~D^{F}_{0}(x_{2} - x_{3})~
[\eta^{\mu_{1}\mu_{3}}~(B_{a_{2}\nu}u_{a_{3}})(x_{1})~C_{a_{1}}^{\nu\mu_{2}}(x_{2}) 
\nonumber\\
- (B_{a_{2}}^{\mu_{3}}u_{a_{3}})(x_{1})~C_{a_{1}}^{\mu_{1}\mu_{2}}(x_{2}) ]
\nonumber\\
+ \delta(x_{1} - x_{3})~\partial^{\mu_{2}}D^{F}_{0}(x_{2} - x_{3})~
(\eta^{\mu_{1}\mu_{3}}~B_{a_{1}\rho}v_{a_{3}}^{\rho} - B_{a_{1}}^{\mu_{3}}v_{a_{3}}^{\mu_{1}})(x_{1})~B_{a_{2}}(x_{2})
\nonumber\\
+ \eta^{\mu_{1}\mu_{3}}~\delta(x_{1} - x_{3})~\partial_{\nu}D^{F}_{0}(x_{1} - x_{2})~
( B_{a_{1}}^{\mu_{2}}u_{a_{3}})(x_{1})~B_{a_{2}}^{\nu}(x_{2}) 
\nonumber\\
- \eta^{\mu_{1}\mu_{3}}~\delta(x_{1} - x_{3})~\partial^{\mu_{2}}D^{F}_{0}(x_{1} - x_{2})~
( B_{a_{1}\nu}u_{a_{3}})(x_{1})~B_{a_{2}}^{\nu}(x_{2}) 
\nonumber\\
-  \eta^{\mu_{1}\mu_{2}}~\delta(x_{1} - x_{3})~\partial_{\nu}D^{F}_{0}(x_{2} - x_{3})~
(B_{a_{1}}^{\mu_{3}}u_{a_{3}})(x_{1})~B_{a_{2}}^{\nu}(x_{2}) 
\nonumber\\
+ \delta(x_{1} - x_{3})~\partial^{\mu_{2}}D^{F}_{0}(x_{2} - x_{3})~
(B_{a_{1}}^{\mu_{3}}u_{a_{3}})(x_{1})~B_{a_{2}}^{\mu_{1}}(x_{2}) \} 
\nonumber\\ 
- (x_{1} \leftrightarrow x_{2}, \mu_{1} \leftrightarrow \mu_{2})
\eea
\bea
T(T^{\mu\nu(2)}(x_{1}),T^{\rho(2)}(x_{2}),T^{(1)}(x_{3}))^{f} = 
\nonumber\\
f_{a_{1}a_{2}a_{3}}~[ \delta(x_{1} - x_{3})~D^{F}_{0}(x_{2} - x_{3})~
(B_{a_{1}\nu}v_{a_{3}}^{\nu})(x_{1})~C_{a_{2}}^{\mu\rho}(x_{2}) 
\nonumber\\
- \eta^{\mu\rho}~\delta(x_{1} - x_{3})~\partial_{\sigma}D^{F}_{0}(x_{2} - x_{3})~
(B_{a_{1}}v_{a_{3}}^{\nu})(x_{1})~B_{a_{2}}^{\sigma}(x_{2})
\nonumber\\
- \delta(x_{1} - x_{3})~\partial^{\rho}D^{F}_{0}(x_{2} - x_{3})~
( B_{a_{1}}v_{a_{3}}^{\mu})(x_{1})~B_{a_{2}}^{\nu}(x_{2}) 
\nonumber\\
+ \delta(x_{2} - x_{3})~\partial^{\mu}D^{F}_{0}(x_{1} - x_{3})~
B_{a_{1}}(x_{1})~(B_{a_{2}}^{\nu}v_{a_{3}}^{\rho})(x_{2}) 
\nonumber\\
+ \eta^{\mu\rho}~\delta(x_{2} - x_{3})~\partial^{\nu}D^{F}_{0}(x_{1} - x_{3})~
B_{a_{1}}(x_{1})~(B_{a_{2}}^{\lambda}v_{a_{3}\lambda})(x_{2}) ]
\nonumber\\
- (\mu \leftrightarrow \nu)
\eea
\bea
T(T^{\mu\nu(2)}(x_{1}),T^{\rho(1)}(x_{2}),T^{(2)}(x_{3}))^{f} = 
\nonumber\\
f_{a_{1}a_{2}a_{3}}~\{ - \delta(x_{1} - x_{2})~D^{F}_{0}(x_{3} - x_{2})~\eta^{\nu\rho}
(B_{a_{1}}u_{a_{2}})(x_{1})~C_{a_{3}}^{\mu}(x_{3}) 
\nonumber\\
+ \delta(x_{1} - x_{2})~\partial_{\lambda}D^{F}_{0}(x_{3} - x_{2})~\eta^{\mu\rho}
[ (B_{a_{1}}u_{a_{2}})(x_{1})~E_{a_{3}}^{\nu\lambda}(x_{3})
+ (B_{a_{1}}v_{a_{2}}^{\nu})(x_{1})~B_{a_{3}}^{\lambda}(x_{3}) ]
\nonumber\\
- \delta(x_{3} - x_{2})~\partial^{\nu}D^{F}_{0}(x_{1} - x_{2})~
B_{a_{1}}(x_{1})~(u_{a_{2}}E_{a_{3}}^{\rho\mu}(x_{3}) \}
\nonumber\\
- (\mu \leftrightarrow \nu)
\eea
\bea
T(T^{\mu\nu(1)}(x_{1}),T^{\rho(2)}(x_{2}),T^{(2)}(x_{3}))^{f} = 
\nonumber\\
f_{a_{1}a_{2}a_{3}}~\{ [ \eta^{\nu\rho}~\delta(x_{2} - x_{1})~\partial_{\lambda}D^{F}_{0}(x_{3} - x_{1})~
(u_{a_{1}}B_{a_{2}}^{\mu})(x_{2})~B_{a_{3}}^{\lambda}(x_{3}) - (\mu \leftrightarrow \nu) ]
\nonumber\\
+ \delta(x_{3} - x_{1})~\partial^{\rho}D^{F}_{0}(x_{2} - x_{1})~
B_{a_{2}}(x_{2})~(u_{a_{1}}E_{a_{3}}^{\mu\nu})(x_{3}) \}
\eea
\end{thm}
Now we can verify gauge invariance for the tree contributions in the third order. We first have
\bea
sT(T^{(2)}(x_{1}),T^{(2)}(x_{2}),T^{(1)}(x_{3}))^{f} = 
\nonumber\\
~[ \delta(x_{1} - x_{3})~D^{F}_{0}(x_{2} - x_{3})~A_{f}(x_{1},x_{2}) 
+ \delta(x_{1} - x_{3})~\partial_{\mu}D^{F}_{0}(x_{2} - x_{3})~B^{\mu}_{f}(x_{1},x_{2})
\nonumber\\
+ \partial_{\mu}\delta(x_{1} - x_{3})~D^{F}_{0}(x_{2} - x_{3})~C^{\mu}_{f}(x_{1},x_{2})
+ \partial_{\mu}\delta(x_{1} - x_{3})~\partial_{\nu}D^{F}_{0}(x_{2} - x_{3})~D^{\mu\nu}_{f}(x_{1},x_{2})
\nonumber\\
+ \delta(x_{1} - x_{3})~\partial_{\mu}\partial_{\nu}D^{F}_{0}(x_{2} - x_{3})~E^{\{\mu\nu\}}_{f}(x_{1},x_{2})]
+ (x_{1} \leftrightarrow x_{2})
\nonumber\\
+ \delta(x_{1} - x_{3})~\delta(x_{2} - x_{3})~F_{f}(x_{3})
\label{221f}
\eea
where the expressions 
$
A,\dots,F
$
can be explicitly computed if we use the first three formulas from the preceding theorem. We only give explicitly
\be
F_{f} = - 2~i~f_{a_{1}a_{2}a_{3}}~v_{a_{1}\mu}E_{a_{2}}^{\mu\nu}B_{a_{3}\nu}.
\label{f}
\ee
Starting from (\ref{2,2,1}) - (\ref{1munu,2rho,2}) we can obtain a similar formula for the canonical contribution
$
sT(T^{(2)}(x_{1}),T^{(2)}(x_{2}),T^{(1)}(x_{3}))^{\rm can};
$
we only notice that 
\be
F^{\rm can} = - F^{f}.
\label{ff}
\ee
In the end we have:
\bea
sT(T^{(2)}(x_{1}),T^{(2)}(x_{2}),T^{(1)}(x_{3})) = 
\nonumber\\
~[ \delta(x_{1} - x_{3})~D^{F}_{0}(x_{2} - x_{3})~A(x_{1},x_{2}) 
+ \delta(x_{1} - x_{3})~\partial_{\mu}D^{F}_{0}(x_{2} - x_{3})~B^{\mu}(x_{1},x_{2})
\nonumber\\
+ \delta(x_{1} - x_{3})~\partial_{\mu}\partial_{\nu}D^{F}_{0}(x_{2} - x_{3})~E^{\{\mu\nu\}}(x_{1},x_{2})]
+ (x_{1} \leftrightarrow x_{2})
\label{221}
\eea
where
\bea
A(x_{1},x_{2}) = i~P_{a}^{\mu}(x_{1})~C_{a\mu}(x_{2})
\nonumber\\
B^{\mu}(x_{1},x_{2}) = i~P_{a\nu}(x_{1})~E_{a}^{\nu\mu}(x_{2}) + i~Q_{a}^{\mu}(x_{1})~D_{a}(x_{2})
\nonumber\\
+ i~Q_{a}^{\mu\nu}(x_{1})~C_{a\nu}(x_{2}) + i~Q_{a}(x_{1})~B_{a}^{\mu}(x_{2})
\nonumber\\
E^{\{\mu\nu\}}_{f}(x_{1},x_{2}) = i~{\cal S}_{\mu\nu}~[{Q_{a}^{\mu}}_{\rho}(x_{1})~E_{a}^{\nu\rho}(x_{2})]
\eea
and we have defined:
\bea
P_{a}^{\mu} \equiv f_{abc}~( C_{b}^{\mu\nu}~v_{c\nu} - C_{b}^{\mu}~u_{c} + B_{b}~\partial^{\mu}\tilde{u}_{c} 
- B_{b\nu}~F_{c}^{\mu\nu} )
\nonumber\\
Q_{a}^{\mu} \equiv f_{abc}~( B_{b}~v_{c}^{\mu} + B_{b}^{\mu}~u_{c})
\nonumber\\
Q_{a}^{\mu\nu} \equiv f_{abc}~( E_{b}^{\nu\mu}~u_{c} + B_{b}^{\mu}~v_{c}^{\nu} - B_{b}^{\nu}~v_{c}^{\mu} )
\nonumber\\
Q_{a} \equiv f_{abc}~\Bigl( - C_{b}^{\rho}~v_{c\rho} - {1\over 2}~E_{b\rho\sigma}~F_{c}^{\rho\sigma} 
+ D_{b}~u_{c} + B_{b}^{\mu}~\partial_{\mu}\tilde{u}_{c} \Bigl).
\label{221ABC}
\eea
However, due to Jacobi identity one can prove that all previous expressions are null. So, in fact we have:
\be
sT(T^{(2)}(x_{1}),T^{(2)}(x_{2}),T^{(1)}(x_{3})) = 0.
\label{221-J}
\ee
From here we have
\bea
sT_{\rm tree}(T(x_{1}),T(x_{2}),T(x_{3})) = 
\nonumber\\
sT(T^{(2)}(x_{1}),T^{(2)}(x_{2}),T^{(1)}(x_{3}))
+ (x_{1} \leftrightarrow x_{3}) + (x_{2} \leftrightarrow x_{3}) = 0
\label{221-tree}
\eea
i.e. there are no anomalies in the tree sector. We remark that (\ref{221-J}) is stronger that (\ref{221-tree}). 
Also, we cannot prove (\ref{221-tree}) only from cohomology considerations because we cannot eliminate
an anomaly of the form
\be
A(x_{1},x_{2},x_{3}) = \delta(x_{1} - x_{3})~\delta(x_{2} - x_{3})~F_{f}(x_{3});
\ee
indeed, if we use Jacobi identity we can prove that such an anomaly is a cocycle i.e. we have
\be
d_{Q}F_{f} = 0
\ee
but it is not a coboundary i.e. it cannot be written in the form
$
A = d_{Q}B - i\partial_{\mu}B^{\mu}
$
so it cannot be eliminated by a redefinition of the chronological products. One must perform an explicit computation
and arrive at (\ref{ff}).

It is natural that (\ref{221}) can be generalized in all other sectors: we have for instance
\bea
sT(T^{(2)}(x_{1}),T^{(2)}(x_{2}),T^{\mu(1)}(x_{3})) = 
\nonumber\\
~[ \delta(x_{1} - x_{3})~D^{F}_{0}(x_{2} - x_{3})~A^{\mu}(x_{1},x_{2}) 
\nonumber\\
+ \delta(x_{1} - x_{3})~\partial_{\nu}D^{F}_{0}(x_{2} - x_{3})~B^{\mu\nu}(x_{1},x_{2}) ]
+ (x_{1} \leftrightarrow x_{2})
\nonumber\\
+ \delta(x_{1} - x_{3})~\delta(x_{2} - x_{3})~F^{\mu}(x_{3})
\label{221mu}
\eea
where
\bea
A^{\mu}(x_{1},x_{2}) = i~R_{a}^{\mu\nu}(x_{1})~C_{a\nu}(x_{2})
\nonumber\\
B^{\mu\nu}(x_{1},x_{2}) = - i~P_{a}^{\mu}(x_{1})~B_{a}^{\nu}(x_{2}) + i~[ Q_{a}^{\nu}(x_{1})~C_{a}^{\mu}(x_{2})
- \eta^{\mu\nu}~Q_{a}^{\rho}(x_{1})~C_{a\rho}(x_{2}) ]
\nonumber\\
+ i~{R_{a}^{\mu}}_{\rho}(x_{1})~E_{a}^{\rho\nu}(x_{2}) + i~\eta^{\mu\nu} S_{a}(x_{1})~D_{a}(x_{2})
\nonumber\\
F^{\mu} = Q_{a\nu}~E_{a}^{\nu\mu}.
\eea
Here we have used the previous notations plus
\bea
R_{a}^{\mu\nu} \equiv - f_{abc}~( C_{b}^{\mu\nu}~u_{c} + B_{b}~F_{c}^{\mu\nu} )
\nonumber\\
S_{a} \equiv f_{abc}~B_{b}~u_{c}.
\eea
These two expressions are also null due to Jacobi identity. We similarly have
\bea
sT(T^{(1)}(x_{1}),T^{(2)}(x_{2}),T^{\mu(2)}(x_{3})) =
\nonumber\\
\delta(x_{2} - x_{1})~D^{F}_{0}(x_{3} - x_{1})~A^{\mu}(x_{2},x_{3}) 
+ \delta(x_{3} - x_{1})~D^{F}_{0}(x_{2} - x_{1})~A^{\prime\mu}(x_{2},x_{3}) 
\nonumber\\
+ \delta(x_{2} - x_{1})~\partial_{\nu}D^{F}_{0}(x_{3} - x_{1})~B^{\mu\nu}(x_{2},x_{3})
+ \delta(x_{3} - x_{1})~\partial_{\nu}D^{F}_{0}(x_{2} - x_{1})~B^{\prime\mu\nu}(x_{2},x_{3})
\nonumber\\
+ \delta(x_{2} - x_{1})~\partial_{\rho}\partial_{\sigma}D^{F}_{0}(x_{3} - x_{1})~E^{\mu\{\rho\sigma\}}(x_{2},x_{3})
\nonumber\\
+ \delta(x_{3} - x_{1})~\partial_{\rho}\partial_{\sigma}D^{F}_{0}(x_{2} - x_{1})~E^{\prime\mu\{\rho\sigma\}}(x_{2},x_{3})
\nonumber\\
+ \delta(x_{2} - x_{1})~\delta(x_{3} - x_{1})~F^{\mu}(x_{1})
\eea
where
\bea
A^{\mu}(x_{2},x_{3}) = i~P_{a\nu}(x_{2})~C_{a}^{\nu\mu}(x_{3})
\nonumber\\
A^{\prime\mu}(x_{2},x_{3}) = i~R_{a}^{\mu\nu}(x_{2})~C_{a\nu}(x_{3})
\nonumber\\
B^{\mu\nu}(x_{2},x_{3}) = i~[ - P_{a}^{\mu}(x_{2})~B_{a}^{\nu}(x_{3}) + \eta^{\mu\nu}~P_{a}^{\rho}(x_{2})~B_{a\rho}(x_{3})
\nonumber\\
- {Q_{a}^{\nu}}_{\rho}(x_{2})~C_{a}^{\rho\mu}(x_{3}) + \eta^{\mu\nu}~Q_{a}(x_{2})~B_{a}(x_{3})
+ Q_{a}^{\nu}(x_{2})~C_{a}^{\mu}(x_{3}) ]
\nonumber\\
B^{\prime\mu\nu}(x_{2},x_{3}) = i~[ R_{a}^{\mu\rho}(x_{3})~{E_{a}^{\nu}}_{\rho}(x_{2}) + Q_{a}^{\nu}(x_{3})~C_{a}^{\mu}(x_{2})
\nonumber\\
- \eta^{\mu\nu}~Q_{a\rho}(x_{3})~C_{a}^{\rho}(x_{2}) - P_{a}^{\mu}(x_{3})~B_{a}^{\nu}(x_{2}) - \eta^{\mu\nu}~S_{a}(x_{3})~D_{a}(x_{2}) ]
\nonumber\\
E^{\mu\{\rho\sigma\}}(x_{2},x_{3}) = i~{\cal S}_{\rho\sigma} [ Q_{a}^{\mu\rho}(x_{2})~B_{a}^{\sigma}(x_{3})
+ \eta^{\mu\rho}~Q_{a}^{\sigma\lambda}(x_{2})~B_{a\lambda}(x_{3}) ]
\nonumber\\
E^{\prime\mu\{\rho\sigma\}}(x_{2},x_{3}) = i~{\cal S}_{\rho\sigma} [ Q_{a}^{\rho}(x_{3})~E_{a}^{\mu\sigma}(x_{2})
- \eta^{\mu\rho}~Q_{a\lambda}(x_{3})~E_{a}^{\lambda\sigma}(x_{2}) ]
\nonumber\\
F^{\mu} = i~f_{a_{1}a_{2}a_{3}}~[ (- v_{a_{1}}^{\mu}B_{a_{2}}^{\rho} + v_{a_{1}}^{\rho}B_{a_{2}}^{\mu})~B_{a_{3}\rho}
+ v_{a_{1}\rho}E_{a_{2}}^{\rho\mu}B_{a_{3}} ].
\eea
This expression is null. Only the proof of
$
F^{\mu} = 0
$
requires some computations; all other expressions have already appeared previously. So we have
\bea
sT_{\rm tree}(T(x_{1}),T(x_{2}),T^{\mu}(x_{3})) = 
sT(T^{(2)}(x_{1}),T^{(2)}(x_{2}),T^{\mu(1)}(x_{3}))
\nonumber\\
+ sT(T^{(1)}(x_{1}),T^{(2)}(x_{2}),T^{\mu(2)}(x_{3})) + (x_{1} \leftrightarrow x_{2}) = 0.
\eea

In the same way we have
\bea
sT(T^{\mu_{1}(2)}(x_{1}),T^{\mu_{2}(2)}(x_{2}),T^{(1)}(x_{3})) =
\nonumber\\
~[ \delta(x_{1} - x_{3})~D^{F}_{0}(x_{2} - x_{3})~A^{\mu_{1}\mu_{2}}(x_{1},x_{2}) 
+ \delta(x_{1} - x_{3})~\partial_{\nu}D^{F}_{0}(x_{2} - x_{3})~B^{\mu_{1}\mu_{2};\nu}(x_{1},x_{2})
\nonumber\\
+ \delta(x_{1} - x_{3})~\partial_{\rho}\partial_{\sigma}D^{F}_{0}(x_{2} - x_{1})~E^{\mu_{1}\mu_{2}\{\rho\sigma\}}(x_{1},x_{2})
\nonumber\\
- (x_{1} \leftrightarrow x_{2}, \mu_{1} \leftrightarrow \mu_{2}) ]
\nonumber\\
+ \delta(x_{1} - x_{3})~\delta(x_{2} - x_{3})~F^{\mu_{1}\mu_{2}}(x_{3})
\eea
which is null because
\bea
A^{\mu_{1}\mu_{2}}(x_{1},x_{2}) = i~R_{a}^{\mu_{1}\nu}(x_{1})~{C_{a}^{\mu_{2}}}_{\nu}(x_{2})
\nonumber\\
B^{\mu_{1}\mu_{2};\nu}(x_{1},x_{2}) = i~[ - \eta^{\mu_{1}\nu}~R_{a}^{\mu_{1}\rho}(x_{1})~B_{a\rho}(x_{2}) 
+ R_{a}^{\mu_{1}\mu_{2}}(x_{1})~B_{a}^{\nu}(x_{2})
\nonumber\\
- \eta^{\mu_{1}\nu}~Q_{a\rho}(x_{1})~C_{a}^{\rho\mu_{2}}(x_{2}) + Q_{a}^{\nu}(x_{1})~C_{a}^{\mu_{1}\mu_{2}}(x_{2})
\nonumber\\
+ \eta^{\mu_{2}\nu}~P_{a}^{\mu_{1}}(x_{1})~B_{a}(x_{2}) + \eta^{\mu_{1}\nu}~S_{a}(x_{1})~C_{a}^{\mu_{2}}(x_{2}) ]
\nonumber\\
E^{\mu\{\rho\sigma\}}(x_{1},x_{2}) = i~{\cal S}_{\rho\sigma} [ \eta^{\mu_{1}\rho}~Q_{a}^{\mu_{2}}(x_{1})~B_{a}^{\sigma}(x_{2})
- \eta^{\mu_{1}\rho}~\eta^{\mu_{2}\sigma}~Q_{a\lambda}(x_{1})~B_{a}^{\lambda}(x_{2})
\nonumber\\
+ \eta^{\mu_{2}\rho}~Q_{a}^{\sigma}(x_{1})~B_{a}^{\mu_{1}}(x_{2})
- \eta^{\mu_{1}\mu_{2}}~Q_{a}^{\rho}(x_{1})~B_{a}^{\sigma}(x_{2}) ]
\nonumber\\
F^{\mu_{1}\mu_{2}} = i~T_{a}^{\mu_{1}} v_{a}^{\mu_{2}} - (\mu_{1} \leftrightarrow \mu_{2}).
\eea
are null. We have introduced
\be
T_{a}^{\mu} \equiv f_{abc}~B_{b} B_{c}^{\mu}.
\ee
Only 
$
T_{a}^{\mu} = 0
$
must be checked. 
\newpage
We also have:
\bea
sT(T^{\mu_{1}(2)}(x_{1}),T^{\mu_{2}(1)}(x_{2}),T^{(2)}(x_{3})) =
\nonumber\\
~[ \delta(x_{3} - x_{2})~D^{F}_{0}(x_{1} - x_{2})~A^{\mu_{1}\mu_{2}}(x_{1},x_{3}) 
+ \delta(x_{3} - x_{2})~\partial_{\nu}D^{F}_{0}(x_{1} - x_{2})~B^{\mu_{1}\mu_{2};\nu}(x_{1},x_{3})
\nonumber\\
+ \delta(x_{1} - x_{2})~\partial_{\nu}D^{F}_{0}(x_{3} - x_{2})~B^{\prime\mu_{1}\mu_{2};\nu}(x_{1},x_{3})
\nonumber\\
+ \delta(x_{3} - x_{2})~\partial_{\rho}\partial_{\sigma}D^{F}_{0}(x_{1} - x_{2})~E^{\mu_{1}\mu_{2}\{\rho\sigma\}}(x_{1},x_{3})
\nonumber\\
+ \delta(x_{1} - x_{2})~\partial_{\rho}\partial_{\sigma}D^{F}_{0}(x_{3} - x_{2})~
E^{\prime\mu_{1}\mu_{2}\{\rho\sigma\}}(x_{1},x_{3})
\nonumber\\
+ \delta(x_{3} - x_{2})~\delta(x_{1} - x_{2})~F^{\mu_{1}\mu_{2}}(x_{2})
\eea
which is null because
\bea
A^{\mu_{1}\mu_{2}}(x_{1},x_{3}) = i~{C_{a}^{\mu_{1}}}_{\nu}(x_{1})~R_{a}^{\mu_{2}\nu}(x_{2})
\nonumber\\
B^{\mu_{1}\mu_{2};\nu}(x_{1},x_{3}) = i~[ - \eta^{\mu_{1}\nu}~B_{a}(x_{1})~P_{a}^{\mu_{2}}(x_{3}) 
- B_{a}^{\nu}(x_{1})~R_{a}^{\mu_{1}\mu_{2}}(x_{3})
\nonumber\\
- \eta^{\mu_{1}\nu}~B_{a\rho}(x_{1})~R_{a}^{\mu_{2}\rho}(x_{3}) + C_{a}^{\mu_{1}\mu_{2}}(x_{1})~Q_{a}^{\nu}(x_{3})
\nonumber\\
- \eta^{\mu_{2}\nu}~C_{a}^{\mu_{2}\rho}(x_{1})~Q_{a\rho}(x_{3}) + \eta^{\mu_{2}\nu}~C_{a}^{\mu_{1}}(x_{1})~S_{a}(x_{3}) ]
\nonumber\\
B^{\prime\mu_{1}\mu_{2};\nu}(x_{1},x_{3}) = i~[ - R_{a}^{\mu_{1}\mu_{2}}(x_{1})~B_{a}^{\nu}(x_{3}) 
+ S_{a}^{\nu}(x_{1})~( - \eta^{\mu_{1}\nu} C_{a}^{\mu_{2}} + \eta^{\mu_{2}\nu}~C_{a}^{\mu_{1}})(x_{3}) ]
\nonumber\\
E^{\mu\{\rho\sigma\}}(x_{1},x_{2}) = i~{\cal S}_{\rho\sigma} 
[ B_{a}^{\rho}(x_{1})~( - \eta^{\mu_{2}\sigma} Q_{a}^{\mu_{1}}
+ \eta^{\mu_{1}\mu_{2}}~Q_{a}^{\rho})(x_{3})
\nonumber\\
+ \eta^{\mu_{1}\rho}~\eta^{\mu_{2}\sigma}~B_{a}^{\lambda}(x_{1})~Q_{a\lambda}(x_{3})
- \eta^{\mu_{1}\rho}~B_{a}^{\mu_{2}}(x_{1})~Q_{a}^{\sigma}(x_{3}) ]
\nonumber\\
E^{\prime\mu\{\rho\sigma\}}(x_{1},x_{2}) = i~{\cal S}_{\rho\sigma} 
[ S_{a}(x_{1})~( - \eta^{\mu_{1}\rho} E_{a}^{\mu_{2}\sigma}
+ \eta^{\mu_{2}\rho}~E_{a}^{\mu_{1}\sigma})(x_{3}) ]
\nonumber\\
F^{\mu_{1}\mu_{2}} = i~( S_{a} E_{a}^{\mu_{2}\mu_{1}} - B_{a}^{\mu_{2}} Q_{a}^{\mu_{1}} 
+ \eta^{\mu_{1}\mu_{2}}`B_{a\rho} Q_{a}^{\rho})
\eea
are null. As a consequence we have
\bea
sT_{\rm tree}(T^{\mu_{1}}(x_{1}),T^{\mu_{2}}(x_{2}),T(x_{3})) =
sT(T^{\mu_{1}(2)}(x_{1}),T^{\mu_{2}(2)}(x_{2}),T^{(1)}(x_{3})) +
\nonumber\\
sT(T^{\mu_{1}(2)}(x_{1}),T^{\mu_{2}(1)}(x_{2}),T^{(2)}(x_{3})) 
-  (x_{1} \leftrightarrow x_{2}, \mu_{1} \leftrightarrow \mu_{2}) = 0.
\eea

Finally we have
\bea
sT(T^{(1)}(x_{1}),T^{(1)}(x_{2}),T^{\mu\nu(2)}(x_{3})) =
\nonumber\\
~[ \delta(x_{1} - x_{3})~\partial_{\rho}D^{F}_{0}(x_{2} - x_{3})~B^{\mu\nu;\rho}(x_{1},x_{2})
\nonumber\\
+ \delta(x_{1} - x_{3})~\partial_{\rho}\partial_{\sigma}D^{F}_{0}(x_{2} - x_{3})~E^{\mu\nu\{\rho\sigma\}}(x_{1},x_{2}) ]
+ (x_{1} \leftrightarrow x_{2})
\nonumber\\
+ \delta(x_{1} - x_{3})~\delta(x_{2} - x_{3})~F^{\mu\nu}(x_{3})
\eea
which is null because
\bea
B^{\mu\nu;\rho}(x_{1},x_{2}) = i~\{ [ \eta^{\mu\rho}~S_{a}(x_{1})~C_{a}^{\nu}(x_{2}) - (\mu \leftrightarrow \nu) ] 
+ R_{a}^{\mu\nu}(x_{1})~B_{a}^{\rho}(x_{2}) \}
\nonumber\\
E^{\mu\nu\{\rho\sigma\}}(x_{1},x_{2}) = i~{\cal S}_{\rho\sigma} [ ( \eta^{\mu\rho} E_{a}^{\nu\sigma}(x_{2}))
- (\mu \leftrightarrow \nu) ]
\nonumber\\
F^{\mu\nu} = i~S_{a} E_{a}^{\mu\nu}. 
\eea
\newpage
We also have
\bea
sT(T^{(1)}(x_{1}),T^{(2)}(x_{2}),T^{\mu\nu(2)}(x_{3})) =
\nonumber\\
\delta(x_{3} - x_{1})~\partial_{\rho}D^{F}_{0}(x_{2} - x_{1})~B^{\mu\nu;\rho}(x_{3},x_{2})
+ \delta(x_{2} - x_{1})~\partial_{\rho}D^{F}_{0}(x_{3} - x_{1})~B^{\prime\mu\nu;\rho}(x_{3},x_{2})
\nonumber\\
+ \delta(x_{3} - x_{1})~\partial_{\rho}\partial_{\sigma}D^{F}_{0}(x_{2} - x_{1})~E^{\mu\nu\{\rho\sigma\}}(x_{3},x_{2})
\nonumber\\
+ \delta(x_{2} - x_{1})~\partial_{\rho}\partial_{\sigma}D^{F}_{0}(x_{3} - x_{1})~E^{\prime\mu\nu\{\rho\sigma\}}(x_{3},x_{2})
\nonumber\\
+ \delta(x_{3} - x_{1})~\delta(x_{2} - x_{1})~F^{\mu\nu}(x_{1})
\eea
which is null because
\bea
B^{\mu\nu;\rho}(x_{3},x_{2}) = i~\{ [ \eta^{\nu\rho}~S_{a}(x_{3})~C_{a}^{\mu}(x_{2}) - (\mu \leftrightarrow \nu) ] 
+ R_{a}^{\mu\nu}(x_{3})~B_{a}^{\rho}(x_{2}) \}
\nonumber\\
B^{\prime\mu\nu;\rho}(x_{3},x_{2}) = i~\{ [ \eta^{\mu\rho}~P_{a}^{\nu}(x_{2})~B_{a}(x_{3}) - (\mu \leftrightarrow \nu) ] 
- Q_{a}^{\rho}(x_{2})~C_{a}^{\mu\nu}(x_{3}) \}
\nonumber\\
E^{\mu\nu\{\rho\sigma\}}(x_{1},x_{2}) = i~{\cal S}_{\rho\sigma} [ \eta^{\mu\rho} S_{a}(x_{3})~E_{a}^{\nu\sigma}(x_{2})
- (\mu \leftrightarrow \nu) ]
\nonumber\\
E^{\prime\mu\nu\{\rho\sigma\}}(x_{1},x_{2}) = i~{\cal S}_{\rho\sigma}[   \eta^{\mu\rho} Q_{a}^{\nu\sigma}(x_{2})
B_{a}(x_{3}) - (\mu \leftrightarrow \nu) ]
\nonumber\\
F^{\mu\nu} = i~T_{a}^{\mu} v_{a}^{\nu} - (\mu \leftrightarrow \nu). 
\eea

It follows that
\bea
sT_{\rm tree}(T(x_{1}),T(x_{2}),T^{\mu\nu}(x_{3})) =
sT(T^{(2)}(x_{1}),T^{(2)}(x_{2}),T^{\mu\nu(1)}(x_{3})) 
\nonumber\\
+ sT(T^{(1)}(x_{1}),T^{(2)}(x_{2}),T^{\mu\nu(1)}(x_{3})) +  (x_{1} \leftrightarrow x_{2}) = 0.
\eea
\section{Conclusions}

We notice in all formulas of the type (\ref{221}) + (\ref{221ABC}) a certain factorization property in two factors emerges;
moreover one of the factors is null because of gauge invariance in the second order. Such a property is lost for the more general case of 
the standard model where one do get anomalies in the third order of the tree sector; the cancelation of this anomaly fixes the Higgs 
coupling \cite{Sc2}. This computation will be addressed in further papers. 

It would be interesting to generalize these ideas in higher orders of perturbation theory.


\begin{thebibliography}{99}

\bibitem{ASD} A. Aste, G. Scharf, M. Duetsch, 
``{\it On gauge invariance and spontaneous symmetry breaking}", 
arXiv: 9705216, J. Phys. A: Math. Gen. {\bf 30} (1997) 5785 - 5792

\bibitem{BS}
N. N. Bogoliubov, D. Shirkov,
``{\it Introduction to the Theory of Quantized Fields}",
John Wiley and Sons, 1976 (3rd edition)

\bibitem{BLOT}
N. N. Bogoliubov, A. A. Logunov, A.I. Oksak, I. Todorov, 
``{\it General Principles of Quantum Field Theory}", Kluwer 1989

\bibitem{D}
M. D\"utsch, ``{\it From Classical Field Theory to Perturbative Quantum Field Theory}", 
Progress in Mathematical Physics {\bf 74}, Springer 2019

\bibitem{DF}
M. D\"utsch, K. Fredenhagen,
``{\it Algebraic Quantum Field Theory, Perturbation Theory,
and the Loop Expansion}",
arXiv: hep-th/0001129, Commun. Math. Phys. {\bf 219} (2001) 5 - 30

\bibitem{DKS} M. Duetsch, F. Krahe, G. Scharf, 
``{\it Scalar QED Revisited}",
Il Nuovo Cimento {\bf 106 A} (3) (1993) 277 - 307

\bibitem{EG}
H. Epstein, V. Glaser,
``{\it The R\^ole of Locality in Perturbation Theory}",
Ann. Inst. H. Poincar\'e {\bf 19 A} (1973) 211-295

\bibitem{Gl}
V. Glaser,
``{\it Electrodynamique Quantique}",
L'enseignement du 3e cycle de la physique en Suisse Romande (CICP), Semestre
d'hiver 1972/73

\bibitem{wick+hopf}
D. R. Grigore,
 ``{\it Wick Theorem and Hopf Algebra Structure in Causal Perturbative Quantum Field Theory}",
arXiv:2202.08056v2 [hep-th]

\bibitem{H}
K. Hepp, ``{\it Renormalization Theory}", in ``{\it Statistical Mechanics and Quantum Field Theory}" pp. 429 - 500,
(Les Houches 1970), C. DeWitt-Morette, Raymond Stora (eds.), Gordon and Breach 1971

\bibitem{P}
J. Polchinski, ``{\it Renormalization and Effective Lagrangians}",
Nucl. Phys. {\bf B 231} (1984) 269 - 295

\bibitem{PS}
G. Popineau, R. Stora, 
``{\it A Pedagogical Remark on the Main Theorem of Perturbative Renormalization Theory}", 
Nuclear Physics {\bf B 912} (2016) 70 - 78

\bibitem{S}
M. Salmhofer, ``{\it Renormalization: An Introduction}", (Theoretical and Mathematical Physics) Springer 1999

\bibitem{Sc1}
G. Scharf,
``{\it Finite Quantum Electrodynamics: The Causal Approach}",
(second edition) Springer, 1995; (third edition) Dover, 2014

\bibitem{Sc2}
G. Scharf,
``{\it Quantum Gauge Theories. A True Ghost Story}",
John Wiley, 2001,
``{\it Quantum Gauge Theories - Spin One and Two}",
Google books, 2010
and
``{\it Gauge Field Theories: Spin One and Spin Two, 100 Years After General Relativity}", Dover 2016

\bibitem{Sto1}
R. Stora,
``{\it Lagrangian Field Theory}",
Les Houches lectures, Gordon and Breach, N.Y., 1971, 
C. De Witt, C. Itzykson eds.

\bibitem{St1}
O. Steinmann,
``{\it Perturbation Expansions in Axiomatic Field Theory}",
Lect. Notes in Phys. {\bf 11}, Springer, 1971

\bibitem{WG}
A. S. Wightman, L. G\aa rding,
``{\it Fields as Operator-Valued Distributions in Relativistic Quantum Field
Theory}", Arkiv Fysik {\bf 28} (1965) 129-184

\end{thebibliography}
\end{document}